\documentclass[apj]{emulateapj}

\usepackage{graphicx}
\usepackage{latexsym}
\usepackage{amsfonts,amsmath,amssymb}
\usepackage[utf8]{inputenc}
\usepackage[T1]{fontenc}
\usepackage{fancyref}
\usepackage[bookmarks]{hyperref}
\usepackage{breakurl}
\usepackage{natbib}
\usepackage{color}
\usepackage{mathtools}
\usepackage{appendix}
\usepackage{rotating}

\definecolor{red}{rgb}{1,0,0}
\newcommand{\red}[1]{\textcolor{red}{#1}}
\renewcommand{\red}[1]{#1}

\hypersetup{colorlinks=false,pdfborder={0 0 0},}

\slugcomment{Submitted to The Astrophysical Journal (ApJ)}

\shorttitle{Transit Light Curves with Finite Integration Time}
\shortauthors{Price \& Rogers}

\begin{document}


\title{Transit Light Curves with Finite Integration Time: Fisher Information
Analysis}


\author{Ellen M. Price}
\affil{California Institute of Technology \\ 1200 East California Boulevard, Pasadena, CA 91125, USA}
\author{Leslie A. Rogers\altaffilmark{1}}
\affil{California Institute of Technology, MC249-17 \\ 1200 East California Boulevard, Pasadena, CA 91125, USA}

\altaffiltext{1}{Hubble Fellow}

\begin{abstract}
\emph{Kepler} has revolutionized the study of transiting planets with its unprecedented photometric precision on more than 150,000 target stars. Most of the transiting planet candidates detected by \emph{Kepler} have been observed as long-cadence targets with 30 minute integration times, and the upcoming Transiting Exoplanet Survey Satellite (TESS) will record full frame images with a similar integration time. Integrations of 30 minutes affect the transit shape, particularly for small planets and in cases of low signal-to-noise. Using the Fisher information matrix technique, we derive analytic approximations for the variances and covariances on the transit parameters obtained from fitting light curve photometry collected with a finite integration time. We find that binning the light curve can significantly increase the uncertainties and covariances on the inferred parameters \red{when comparing scenarios with constant total signal-to-noise (constant total integration time in the absence of read noise)}. Uncertainties on the transit ingress/egress time increase by a factor of 34 for Earth-size planets and 3.4 for Jupiter-size planets around Sun-like stars for integration times of 30 minutes compared to instantaneously-sampled light curves. Similarly, uncertainties on the mid-transit time for Earth and Jupiter-size planets increase by factors of 3.9 and 1.4. Uncertainties on the transit depth are largely unaffected by finite integration times. While correlations among the transit depth, ingress duration, and transit duration all increase in magnitude with longer integration times, the mid-transit time remains uncorrelated with the other parameters. We provide code \red{in Python and \textit{Mathematica}} for predicting the variances and covariances at \url{www.its.caltech.edu/~eprice}.
\end{abstract}


\keywords{methods: analytical --- occultations --- planetary systems --- planets and satellites: general --- techniques: photometric}

\bibliographystyle{apj}

\section{Introduction}
The {\it Kepler} mission has discovered thousands of transiting planet candidates, ushering in a new era of exoplanet discovery and statistical analysis. The light curve produced by the transit of a planet across the disk of its star can provide insights into the planet inclination; eccentricity; stellar density; multiplicity, using transit-timing variations (TTVs); and planet atmosphere, using transmission spectroscopy. As the analysis of {\it Kepler} data pushes toward Earth-size planets on Earth-like orbits, it is imperative to account for and to understand the uncertainties and covariances in the parameters that can be inferred from a transit light curve.

\citet[][hereafter C08]{CarterEt2008ApJ} performed a Fisher information analysis on a simplified trapezoidal transit light curve model to derive analytic approximations for transit parameters as well as their uncertainties and covariances. These analytic approximations are useful when planning observations (e.g. assessing how many transits are needed for a given signal-to-noise on the derived planet properties), optimizing transit data analysis (e.g. by choosing uncorrelated combinations of parameters), and estimating the observability of subtle transit effects. However, C08 assumed that the light curves were instantaneously sampled, and as a result did not account for the effect of finite integration times.

Most {\it Kepler} planets are observed with long-cadence, 30-minute integration times. A finite integration time (temporal binning) induces morphological distortions in the transit light curve. \citet{Kipping2010MNRAS} studied these distortions and their effect on the measured light curve parameters. The main effect of finite integration time is to smear out the transit light curve into a broader shape, with the apparent ingress/egress duration increased by an integration time, and the apparent duration of the flat bottom of totality is decreased by an integration time. As a consequence, the retrieved impact parameter may be overestimated, while the retrieved stellar density is underestimated. \citet{Kipping2010MNRAS} provides approximate analytic expressions for the effect of integration on the light curves and discusses numerical integration techniques to compensate for these effects, \red{but his purpose was not to} undertake a full Fisher analysis or to study the covariances between various parameters induced by the finite integration time.

In this paper, we extend the analysis of \citet{CarterEt2008ApJ} to account for the effects of a finite integration time. We apply a Fisher information analysis to a time-integrated trapezoidal light curve to derive analytic expressions for the uncertainties and covariances of model parameters derived from fitting the light curve (Section~\ref{sec:FI} and Appendix~\ref{adx:fullFI}). We verify these expressions with \red{Markov chain Monte Carlo fits to synthesized {\it Kepler} long cadence data} (Section~\ref{sec:MCMC}). Our analytic expressions can readily be substituted for those of \citet{CarterEt2008ApJ} (e.g. their Equation 31) when calculating the variances of transit parameters for any integration time. We provide code online at \url{www.its.caltech.edu/~eprice} \red{that calculates the estimated variances and covariances of the trapezoidal light curve parameters for any set of system parameters}. We discuss and conclude in Sections~\ref{sec:dis} and \ref{sec:con}.

\section{Linear approximation to binned transit light curve}
A transit light curve represents the flux, as a function of time, received from a star as a planet eclipses it. In general, modeling the transit light curve involves three main ingredients. First, there is some model or parameterization of spatial variations in the surface brightness of the star (due to limb darkening and/or star spots). Second, the stellar flux received is calculated as a function of the planet-star center-to-center sky-projected distance \citep{Mandel&Agol2002ApJ,SeagerMO2003ApJ}. Third, the planet-star center-to-center sky-projected distance must be evaluated as a function of time, either using two-body Keplerian motion, or through {\it N}-body simulations if there are multiple dynamically interacting planets.

\begin{figure}
{\includegraphics[width=\columnwidth]{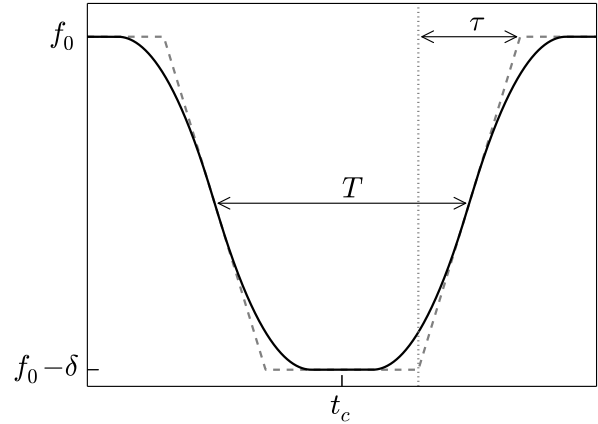}}
\caption{\label{fig:lightcurve} \red{Comparing the linear trapezoidal transit model (time vs. flux) of C08 (dashed, gray line) to the binned trapezoidal model of this work (solid, black line), for an illustrative set of parameters. For both models, $T$ is the full-width half-max duration of the transit event, $\tau$ is the duration of ingress/egress, $\delta$ is the transit depth, $f_0$ is the out-of-transit flux level, and $t_c$ is the time of transit center.}}
\end{figure}

Following C08, we consider a simplified model for the light curve of a dark spherical planet of radius $R_p$ transiting in front of a spherical star  of radius $R_*$. We neglect limb darkening and assume that the star has a uniform surface brightness $f_0$. We assume that the orbital period of the planet is long compared to the transit duration, so that the motion of the planet can be approximated by a constant velocity across the stellar disk. We then adopt the C08 light curve model that approximates the transit light curve as a piecewise linear function in time (Equation~\ref{eqn:lclinear}), \red{where $\delta$ is the transit depth, $T$ is the full-width half-max transit duration, and $\tau$ is the ingress/egress duration, as shown in Figure~\ref{fig:lightcurve}; see \citet{Winn2011} for a more complete description of these parameters.}

\begin{multline}
F_l \left( t;~ t_c, \delta, \tau, T, f_0 \right) = \\
	\begin{cases}
		f_0 - \delta, &{}
			\left| t - t_c \right| \le \frac{T}{2} - \frac{\tau}{2} \\
		\!\begin{aligned}
		    & f_0 - \delta + \frac{\delta}{\tau} \\
		    & ~\times \left( \left| t - t_c \right| - \frac{T}{2} + \frac{\tau}{2} \right),
		\end{aligned} &
			\frac{T}{2} - \frac{\tau}{2} < \left| t - t_c \right| <
			\frac{T}{2} + \frac{\tau}{2} \\
		f_0, &
			\left| t - t_c \right| \ge \frac{T}{2} + \frac{\tau}{2}
	\end{cases}
	\label{eqn:lclinear}
\end{multline}

As in C08, the parameters of the linear trapezoidal light curve model are related to the physical properties of the system (semi-major axis $a$, inclination $i$, eccentricity $e$, longitude of periastron $\omega$, and mean motion $n$) by,
\begin{eqnarray}
\delta &=& f_0r^2 = f_0\left(\frac{R_p}{R_*}\right)^2 \label{eq:r} \\
T &=& 2\tau_0\sqrt{1-b^2} \label{eq:T} \\
\tau &=& 2\tau_0\frac{r}{\sqrt{1-b^2}},\label{eq:tau}
\end{eqnarray}

\noindent where

\begin{eqnarray}
b &\equiv& \frac{a\cos{i}}{R_*}\left(\frac{1-e^2}{1+e\sin\omega}\right)\\
\tau_0 &\equiv& \frac{R_*}{an}\left(\frac{\sqrt{1-e^2}}{1+e\sin\omega}\right).
\end{eqnarray}

\noindent Here, $b$ is the impact parameter, and $\tau_0$ is the timescale for the planet to move one stellar radius (projected on the sky).

We integrate the C08 linear transit light curve in time, to account for a finite integration time, $\mathcal{I}$.
We denote by $F_{lb}(t)$ the average received flux (in the linear model) over a time interval $\mathcal{I}$ centered on time $t$. We restrict our consideration to scenarios with $\mathcal{I}<T-\tau$, because otherwise the measurement of the transit depth during totality will be completely washed out by the integration time.

{\red{
\begin{multline}
F_{lb} \left( t;~ t_c, \delta, \tau, T, f_0, \mathcal{I} \right) = \\
	\begin{cases}
	    f_0 - \delta, & \mathrm{if~} \left| t - t_c \right| \le \frac{T}{2} - \frac{\tau}{2} - \frac{\mathcal{I}}{2} \\[0.5em]
	    \makebox[0pt][l]{$f_0 - \delta + \frac{\delta}{2\tau \mathcal{I}} \left( \left| t - t_c \right| + \frac{\mathcal{I}}{2} - \frac{T}{2} + \frac{\tau}{2} \right)^2$,} \\[0.5em]
		& \mathrm{if~} \frac{T}{2} - \frac{\tau}{2} - \frac{\mathcal{I}}{2} <
		    \left| t - t_c \right| \le{} \frac{T}{2} - \left| \frac{\tau}{2} -
		    \frac{\mathcal{I}}{2}\right| \\[0.5em]
		\makebox[0pt][l]{$f_0 - \delta + \frac{\delta}{\max\left({\tau, \mathcal{I}}\right)} \left( \left| t - t_c \right| - \frac{T}{2} + \frac{\max\left({\tau, \mathcal{I}}\right)}{2} \right)$,} \\[0.5em]
		& \mathrm{if~} \frac{T}{2} - \left| \frac{\tau}{2} - \frac{\mathcal{I}}{2}\right|  < \left| t - t_c \right| < \frac{T}{2} + \left| \frac{\tau}{2} - \frac{\mathcal{I}}{2}\right|   \\[0.5em]
		\makebox[0pt][l]{$f_0 - \frac{\delta}{2 \tau \mathcal{I}} \left( \frac{T}{2} + \frac{\tau}{2} + \frac{\mathcal{I}}{2} - \left| t - t_c \right| \right)^2$,} \\[0.5em]
		& \mathrm{if~} \frac{T}{2} + \left| \frac{\tau}{2} - \frac{\mathcal{I}}{2}\right|   \le \left| t - t_c \right| < \frac{T}{2} + \frac{\tau}{2} + \frac{\mathcal{I}}{2} \\[0.5em]
		f_0, & \mathrm{if~} \left| t - t_c \right| \ge \frac{T}{2} + \frac{\tau}{2} + \frac{\mathcal{I}}{2}
	\end{cases}
	\label{eqn:lcbinned}
\end{multline}
}}

\noindent {\red{Equation~\ref{eqn:lcbinned} gives the the binned light curve model for cases where $\mathcal{I}<T-\tau$. Moving forward, we will differentiate between scenarios where the integration times less than the ingress/egress time, $\mathcal{I} < \tau$ (case 1), and and scenarios where $\mathcal{I} > \tau$ (case 2).
}}

\section{Fisher information analysis}
\label{sec:FI}
When fitting a model to observed data, one is evaluating the likelihood of the observed data, $\{y\}$, conditioned on a hypothesis, typically given in the form of a parametric model $f(\{a\})$. In this scenario, one is asking the question, ``What is the probability of each of my data given the set of parameters $\{a\}$?'' In addition to seeking the parameters that maximize the likelihood of observing the data, one is often also interested in the sensitivity of the data to the model parameters, with the aim of placing confidence intervals on the ``best-fitting'' parameters. In this case one is asking, ``What is the sensitivity of my data to small changes in the model parameters?''

The Fisher information formalism provides a means of addressing this question. The diagonal elements of the Fisher information matrix encode the variance of each parameter, and the off-diagonal elements give the covariances of the parameters. The magnitudes of the variances and covariances are a function of both the nature of the model and the uncertainties in the data.

In the special case that the observed data are normally, identically, and independently distributed about the model \red{with constant total uncertainty $\sigma$}, and assuming flat priors on each of the model parameters, the Fisher information matrix is simply the inverse of the covariance matrix that is often a byproduct of a traditional least-squares analysis. The full details of the Fisher information matrix derivation is given in Appendix~\ref{adx:fullFI}.

We apply Fisher information formalism to the integrated trapezoidal light curve model to derive two different sets of covariance matrices. First, we use the $\{ t_c, \tau, T, \delta, f_0 \}$ parametrization in both the $\tau > \mathcal{I}$ and $\tau < \mathcal{I}$ regimes; then, we transform these matrices to a second, more physical parametrization adopted by C08, $\{ t_c, b^2, \tau_0^2, r, f_0 \}$. \red{We consider this parameterization to be more physical because its parameters are more closely related to the properties of the system, and they are of more astronomical interest when characterizing planet systems.}

All of the variances and covariances we derive in Appendix~\ref{adx:fullFI} are scaled by the common factor $\sigma^2 / \Gamma$, where $\Gamma$ is the sampling rate of the light curve. \red{Formally, the analysis assumes a single transit light curve sampled at rate $\Gamma$.} For phase-folded data spanning several transits, we can let $T_\mathrm{tot} = P$, where $T_\mathrm{tot}$ is the length of the observation baseline and $P$ is the orbital period; then we may substitute an ``effective'' sampling rate, $\Gamma_\mathrm{eff}$, which can be at most $N \Gamma$ if $N$ transits were observed. Practically, we can define $\Gamma_\mathrm{eff}$ as the reciprocal of the average time between consecutive phase-folded time points (see Section~\ref{sec:PhaseSampling} for a discussion of the effects of phase sampling on $\Gamma_\mathrm{eff}$). $\Gamma_\mathrm{eff}$ can be expressed independently of the integration time $\mathcal{I}$ and can be substituted directly for $\Gamma$ when appropriate. \red{The Fisher information analysis assumes that the orbital period $P$ is known with absolute certainty when applied to phase-folded data, so the uncertainty on the transit midtime $t_c$ is not representative of that for individual transits if $\Gamma_\mathrm{eff}$ is used.}

In some cases the out-of-transit flux level, $f_0$, is known to high enough precision that it can be fixed in the fitting process. In the following sections, we assume this is the case and look at the implications of Equations~\ref{eqn:cov1}, \ref{eqn:cov2}, \ref{eqn:cov3}, and \ref{eqn:cov4} for the precision of the transit parameters derived from fitting flux-normalized transits $\left(f_0=1\right)$.

It turns out that, under the assumption of a multivariate normal distribution of the parameters, marginalizing over $f_0$ is equivalent to removing the row and column that contain the variance of $f_0$ and covariances of $f_0$ with the other parameters and substituting the mean value of $f_0$ (here assumed to be $f_0=1$) in the remaining matrix \citep[e.g.,][]{Coe2009arXiv}.

\section{\red{Numerical experiments}}
\label{sec:MCMC}
The covariance expressions (Equations~\ref{eqn:cov1}, \ref{eqn:cov2}, \ref{eqn:cov3}, and \ref{eqn:cov4}) predict the uncertainties in the model parameters for a given dataset; we investigated their applicability with numerical experiments, in which we generated simulated transit photometry data and then fit the light curves to retrieve the transit parameters, uncertainties, and covariances.

We synthesized light curves by numerically integrating the C08 linear flux model over a 30-minute integration time, with time steps evenly spaced at 3-minute intervals. This choice corresponds to an effective sampling rate $\Gamma_\mathrm{eff} = 10\Gamma$, with $\Gamma$ the 30-minute {\it Kepler} long-cadence rate. We assumed that ten transits were measured at sampling rate $\Gamma$ and then phase-folded over one orbital period (see Section~\ref{sec:PhaseSampling} for a discussion on the effects of sampling rate and phase). We then added white noise using a pseudo-random number generator to the relative photometry at a level of $\sigma_i = 5 \times 10^{-5}$ per long-cadence sample.

To retrieve the variances (uncertainties) and covariances of the transit parameters, we fit a binned trapezoidal light curve model {\red{(Equation~\ref{eqn:lcbinned})}} to our simulated, phase-folded photometry. We used the known, ``true'' light curve parameters as a starting point to sample the \red{joint four-dimensional likelihood distribution} with {\tt emcee}, an affine-invariant ensemble sampler for Markov chain Monte Carlo (MCMC) implemented in the Python programming language \citep[][proposed by \citealt{goodman2010ensemble}]{ForemanMackeyEt2013PASP}. The burn-in of the MCMC was sufficiently long that the starting parameters should not have impacted the resulting posteriors. Using $3 \times 10^4$ MCMC chain samples, we estimated the covariance matrix with the Python \texttt{numpy.cov} method.

For our numerical experiments, we considered nominal planet-star parameters corresponding to a Solar-twin ($R_\star = R_{\odot}$, $M_\star = M_{\odot}$) transited by a Jupiter-sized planet ($R_p / R_\star = 0.1$) on an eccentric orbit ($e = 0.16$) transiting at periastron ($\omega = \pi/2$) with $P = 9.55~\rm{days}$ at impact parameter $b = 0.2$. We explored the effect of varying the parameters $R_p / R_\star$, $e$, $b$, and $P$ in Figures \ref{fig:varRpTrapz} to \ref{fig:varPeriodTrapz}. \red{For the experiment varying $P$ only, we fixed the baseline of the observations to be $90$ days instead of assuming that ten transits were observed.} Our analytic expressions for the variances and covariances of both the shape parameters and the physical parameters agree well with the results of the MCMC numerical experiments.

\section{Results}
\begin{figure*}
\begin{center}
\includegraphics[width=\linewidth]{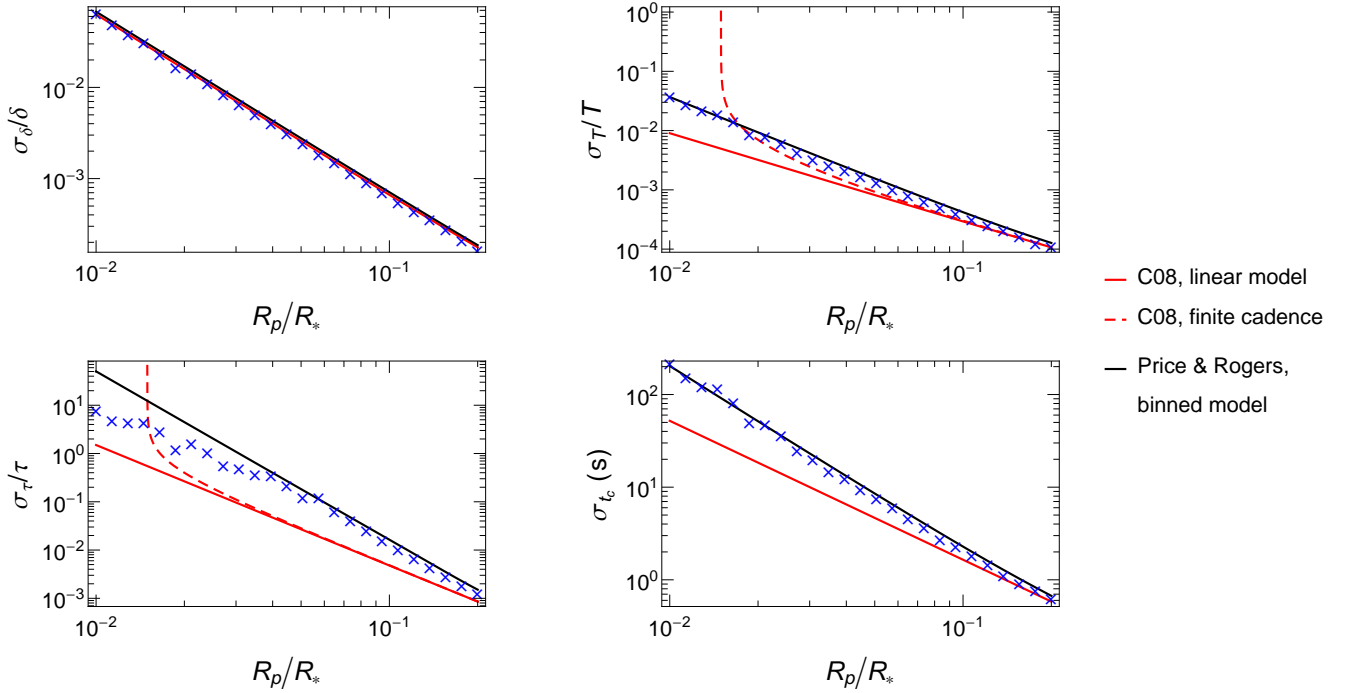}
\caption{\label{fig:varRpTrapz}
Relative uncertainties of the trapezoidal transit parameters derived from \textit{Kepler} long-cadence data, as a function of $R_p/R_\star$. The fiducial planet and star properties assumed are: $R_\star = R_{\odot}$, $M_\star = M_{\odot}$, $e = 0.16$, $P=9.55~\rm{days}$, and $b=0.2$. The solid red line gives $R_p/R_\star$, corresponding to  the analytic predictions from \citet{CarterEt2008ApJ} (their Equation 20), the dashed red curve gives the analytic predictions from \citet{CarterEt2008ApJ} including a finite-cadence correction (their Equation 26), and the solid black curve presents the analytic predictions accounting for a finite integration time from this work (Equations \ref{eqn:cov1} and \ref{eqn:cov2}). The uncertainties derived from an MCMC analysis of simulated long-cadence Kepler data (blue crosses) agree well with the predictions of this work; we plot the measured uncertainty scaled by the true value of the parameter (where appropriate), so this plot does not reflect any systematic error in parameter measurement.}
\end{center}
\end{figure*}

One of the most important effects of a finite integration time is to increase the magnitudes of the covariances among the parameters derived from fitting a transit light curve. A finite integration time also increases the variances of both the shape and the physical parameters derived from a transit light curve. In some regimes (small $R_p/R_\star$, short $P$, and low signal-to-noise ratio, S/N) the variances on $T$ and $\tau$ can increase by an order of magnitude (Figure~\ref{fig:varRpTrapz}). The scaling of the variances with $R_p/R_\star$ and $P$ is also affected. Finite integration time makes the scaling of the variances with $R_p/R_\star$ universally stronger, while the dependence on $P$ becomes more complicated than a simple power law relation; the orbital period of a planet influences whether it falls in the $\tau <\mathcal{I}$ or $\mathcal{I}<\tau$ regime, and the variance on $\tau$ increases substantially once $\tau <\mathcal{I}$.

The variance on the transit depth, $\delta$, which governs the precision with with which the planet-to-star radius ratio can be inferred, is not strongly affected by a finite integration time. Accounting for a finite integration time, the variance in $\delta$ increases by a factor 1.14 \red{(assuming the nominal orbit parameters described in the previous section)}. This corresponds to a factor 1.06 increase on the uncertainty in the transit depth for {\it Kepler}'s 30-minute long-cadence data compared to the 1-minute short-cadence data. The transit depth shows stronger correlations with $T$ (negative correlation) and $\tau$ (positive correlation), especially in cases with low S/N.

The time of transit center is especially important for measuring transit timing variations (TTVs).
The precision that can be obtained with the time of transit center $t_c$ scales with the signal-to-noise of the transit detection in the light curve. From Equations~\ref{eqn:cov1} and \ref{eqn:cov2},

\begin{displaymath}
\sigma_{t_c} = \left\{
     \begin{array}{lc}
       \frac{1}{Q}\sqrt{\frac{\tau T}{2}} \frac{1}{\sqrt{1-\frac{\mathcal{I}}{3\tau}}} &  \tau\geq \mathcal{I}\\
       \frac{1}{Q}\sqrt{\frac{\mathcal{I} T}{2}} \frac{1}{\sqrt{1-\frac{\tau}{3\mathcal{I}}}} &  \mathcal{I}>\tau
     \end{array}
   \right.,
\end{displaymath}

\noindent where $Q = \sqrt{\Gamma T}\frac{\delta}{\sigma}$ is the total signal-to-noise ratio of the transit in the limit $r\to0$. We note that for TTVs, $t_c$ is measured for each individual transit; thus, the single transit sampling rate, $\Gamma$, should be used to predict $\sigma_{t_c}$ for an individual transit, and not $\Gamma_\mathrm{eff}$ for a phase-folded transit. From C08, the expected uncertainty in the transit time derived from an instantaneously sampled transit light curve is

\red{
\begin{equation}
\lim_{\mathcal{I} \to 0} \sigma_{t_c} = \frac{1}{Q}\sqrt{\frac{\tau T}{2}}.
\end{equation}
}

\noindent A finite integration time introduces a $\mathcal{I}/\tau$ dependent correction factor, and effectively substitutes $\mathcal{I}$ for $\tau$ in the formula for $\sigma_{t_c}$ in the $\mathcal{I}>\tau$ regime. Importantly, $t_c$ remains uncorrelated with the other parameters when a finite integration time is taken into account.

\begin{figure*}
\begin{center}
\includegraphics[width=\linewidth]{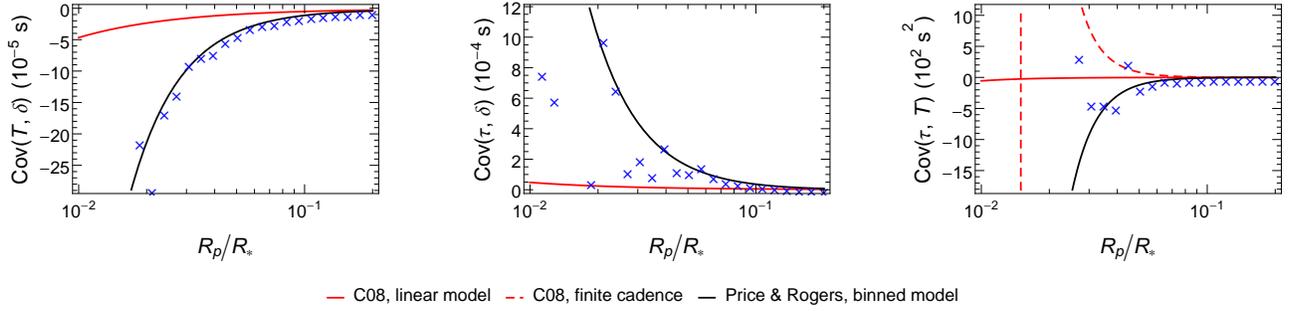}
\caption{\label{fig:covRpTrapz}
Covariances in the trapezoidal approximation transit parameters derived from simulated \textit{Kepler} long-cadence data, as a function of $R_p/R_\star$, corresponding to the same scenarios presented in Figure~\ref{fig:varRpTrapz}.}
\end{center}
\end{figure*}

The dependence of variances and covariances of the light curve parameters on $R_p/R_\star$ are shown in Figures~\ref{fig:varRpTrapz} and \ref{fig:covRpTrapz}, respectively. The MCMC variances and covariances start to deviate from the analytic predictions once $R_p/R_\star < 0.04$. This could be due to the fact that our integral approximation to the finite sums is breaking down at that point. Indeed, we see that $\Gamma_\mathrm{eff}\;\tau < 3$ for those small planet radii (see Section~\ref{sec:PhaseSampling}). Another possibility is that the posterior distribution of $\tau$ is no longer Gaussian at this point (see Section~\ref{sec:NonGaussian}).

\begin{figure*}
\begin{center}
\includegraphics[width=\linewidth]{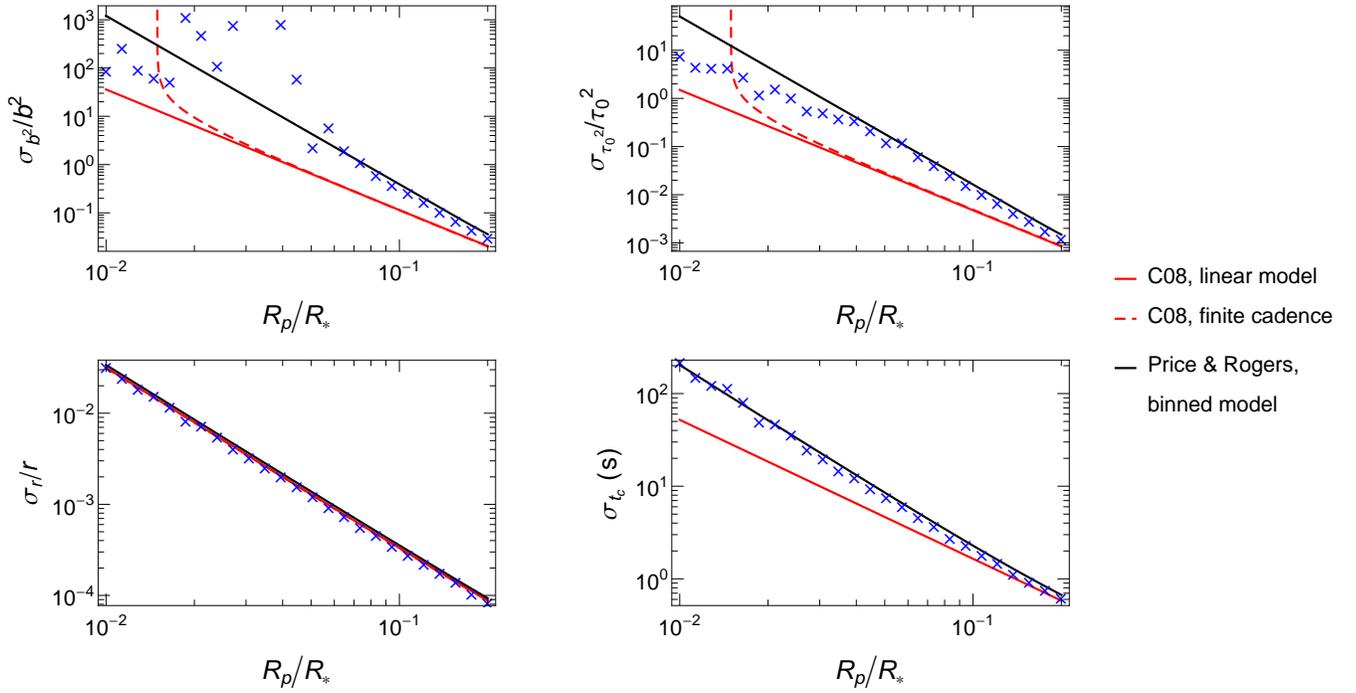}
\caption{\label{fig:varRpPhys}
Relative uncertainties in the physical transit parameters derived from simulated \textit{Kepler} long-cadence data, as a function of $R_p/R_\star$, corresponding to the same scenarios presented in Figure~\ref{fig:varRpTrapz}. We scale by the true value of the parameter to isolate the uncertainty from systematic error in parameter measurement.}
\end{center}
\end{figure*}

In Figure~\ref{fig:varRpPhys}, we plot the predicted and measured uncertainties of the ``physical'' parameters $r$, $b^2$, and $\tau_0^2$. The deviation of the relative uncertainty in $\tau_0^2$ at $R_p/R_\star < 0.04$ seems to be caused by the corresponding deviation of the relative uncertainty of $\tau$. We note a significant deviation in the measured $\sigma_{b^2}$, but it does not have such an obvious explanation. Our results illustrate that $b^2$ is the most difficult of the physical parameters to constrain, particularly at small $R_p/R_\star$. \red{As an additional test, by sampling from these parameters, instead of the trapezoidal parameters, in an MCMC fit to synthetic light curve data, we verified that the variances and covariances of the physical parameters obtained from fitting are consistent with these results.}

\begin{figure*}
\begin{center}
\includegraphics[width=\linewidth]{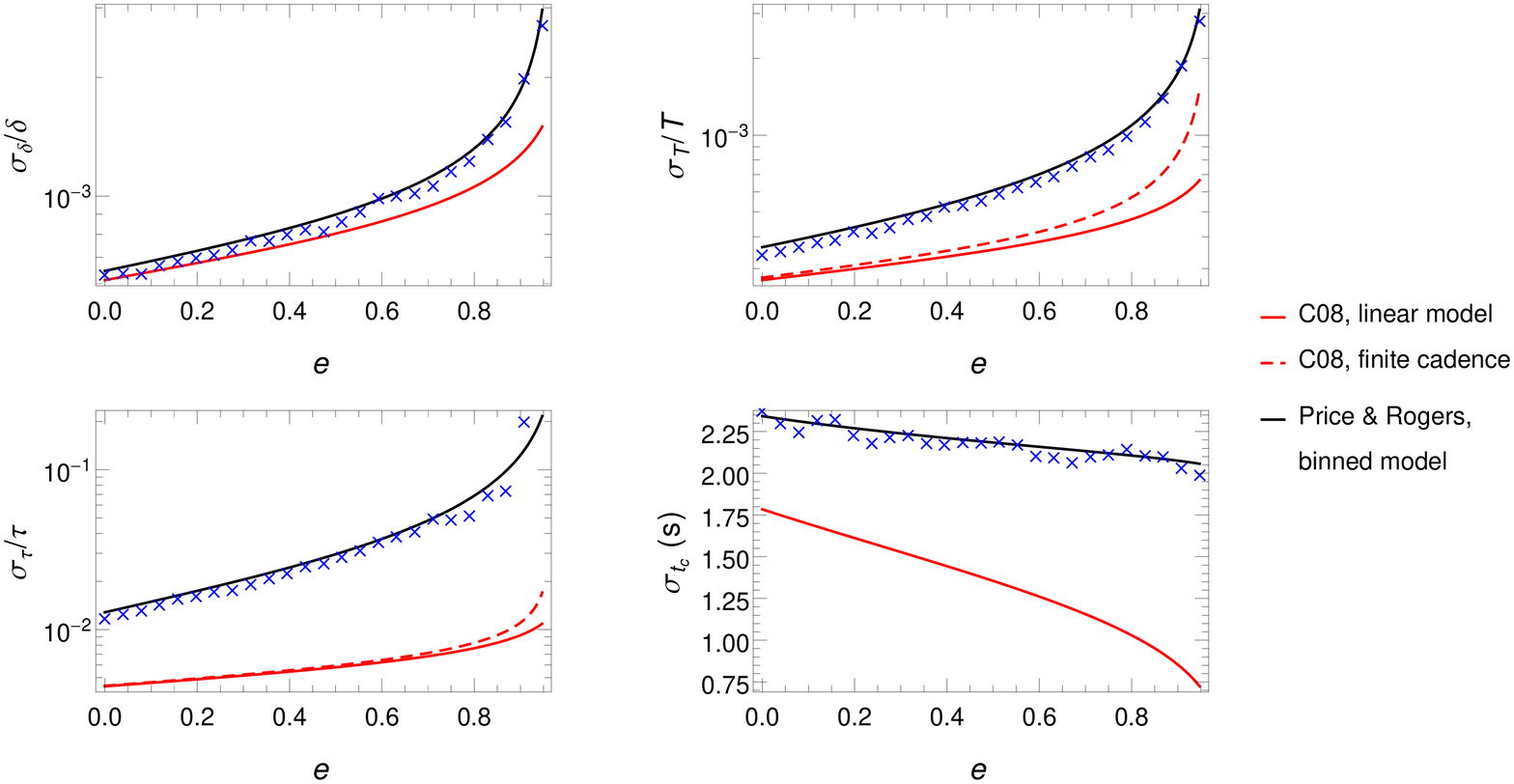}
\caption{\label{fig:varEccTrapz}
Relative uncertainties of the trapezoidal transit parameters derived from \textit{Kepler} long-cadence data, as functions of eccentricity $e$. We have assumed nominal planet parameters $R_\star = R_\odot$, $M_\star = M_\odot$, $P = 9.55~\text{days}$, $b = 0.2$, and $r = 0.1$. To isolate systematic error in parameter measurement from parameter uncertainty, the relative uncertainties are scaled by the true value of the parameter.}
\end{center}
\end{figure*}

\begin{figure*}
\begin{center}
\includegraphics[width=\linewidth]{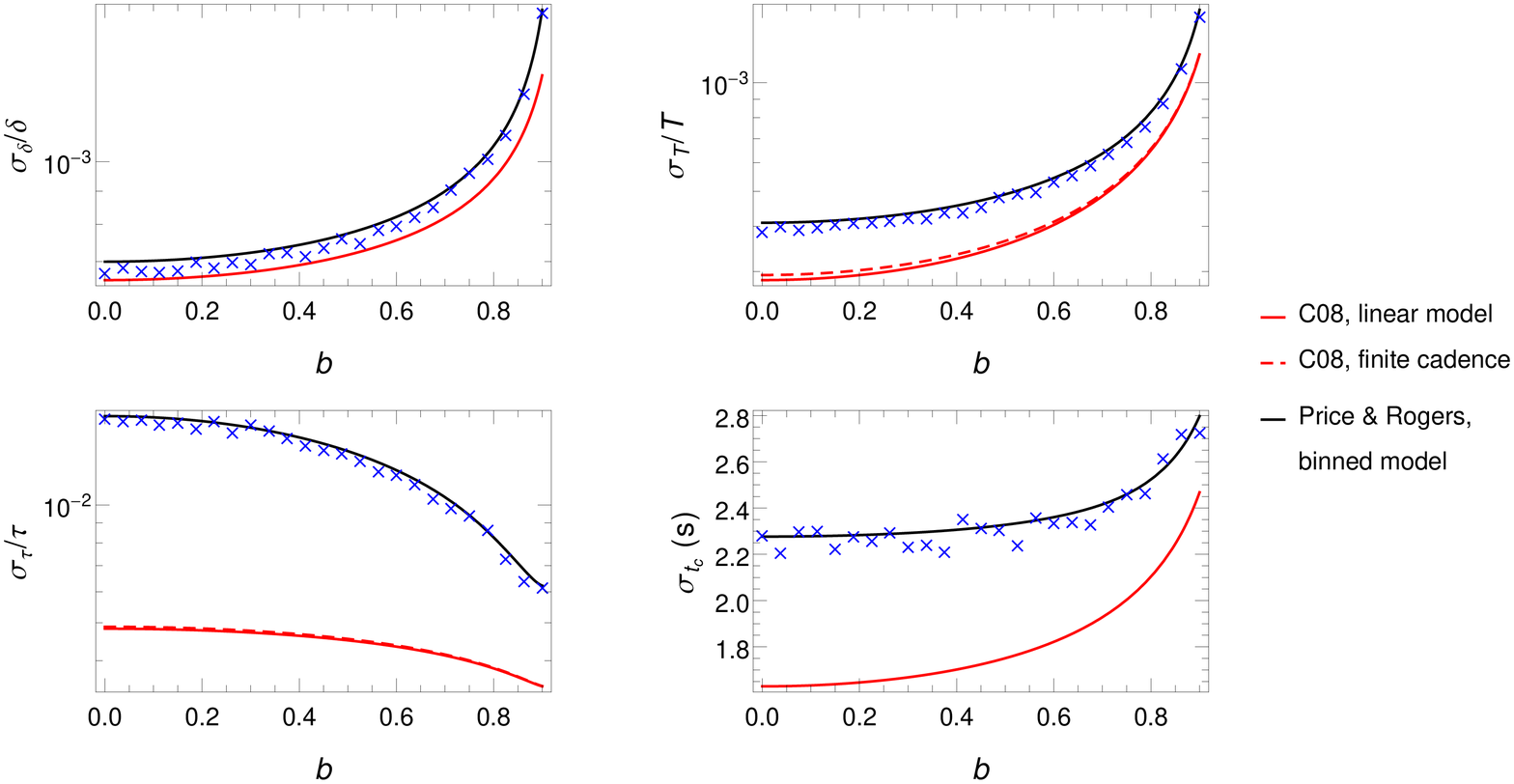}
\caption{\label{fig:varImpactTrapz}
Relative uncertainties of the trapezoidal transit parameters derived from \textit{Kepler} long-cadence data, as functions of impact parameter $b$. We have assumed nominal planet parameters $R_\star = R_\odot$, $M_\star = M_\odot$, $P = 9.55~\text{days}$, $e = 0.16$, and $r = 0.1$. To isolate systematic error in parameter measurement from parameter uncertainty, the relative uncertainties are scaled by the true value of the parameter.}
\end{center}
\end{figure*}

\begin{figure*}
\begin{center}
\includegraphics[width=\linewidth]{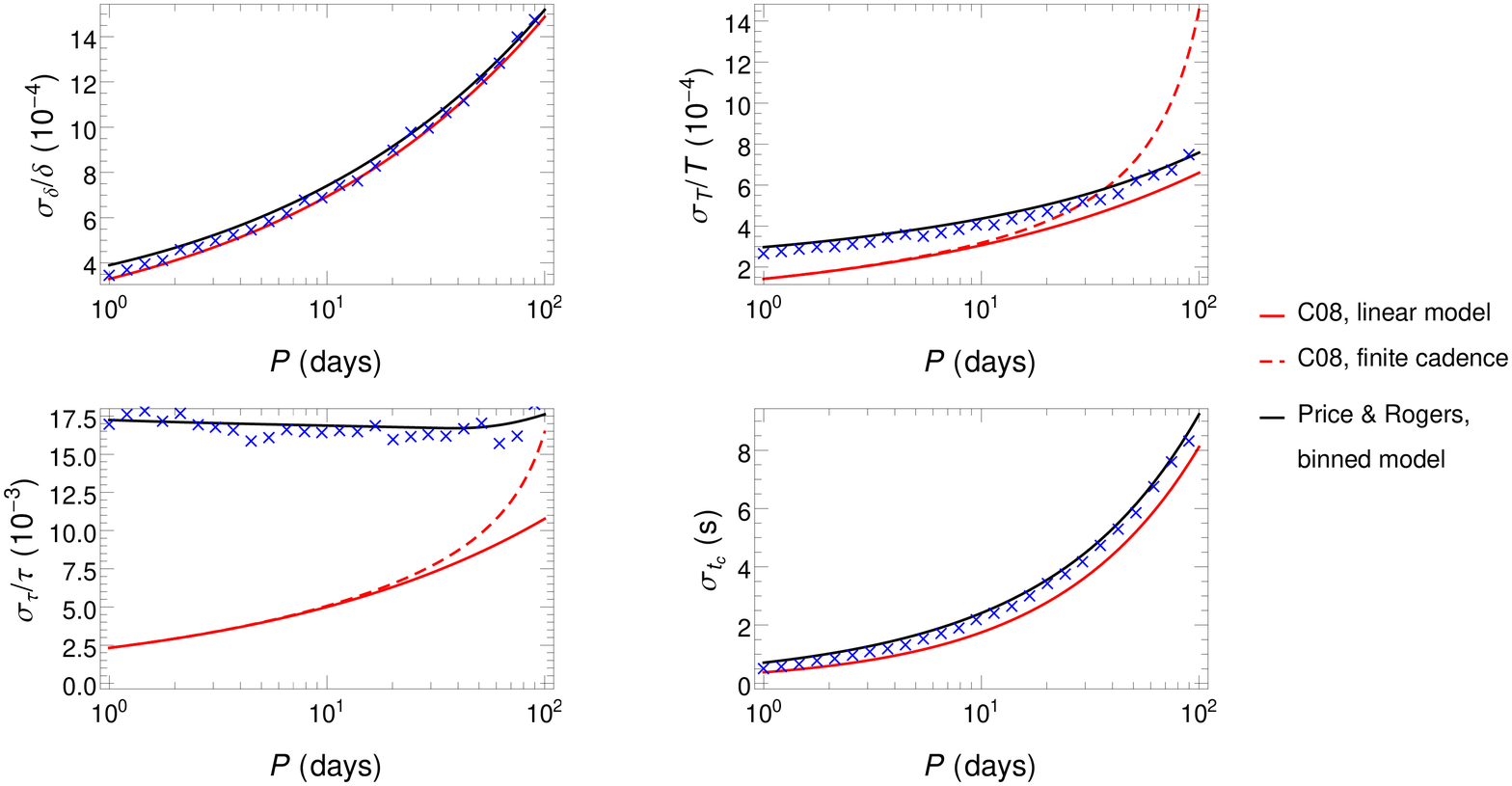}
\caption{\label{fig:varPeriodTrapz}
\red{Relative uncertainties of the trapezoidal transit parameters derived from \textit{Kepler} long-cadence data, as functions of impact parameter $P$; for this experiment only, we fix the baseline of the observations to be $90$ days, instead of assuming that ten transits were observed. We have assumed nominal planet parameters $R_\star = R_\odot$, $M_\star = M_\odot$, $b = 0.2$, $e = 0.16$, and $r = 0.1$. To isolate systematic error in parameter measurement from parameter uncertainty, the relative uncertainties are scaled by the true value of the parameter.}}
\end{center}
\end{figure*}

Figures~\ref{fig:varEccTrapz}, \ref{fig:varImpactTrapz}, and \ref{fig:varPeriodTrapz} show the predicted and measured uncertainties of the trapezoidal light curve parameters as functions of eccentricity $e$, impact parameter $b$, and orbital period $P$, respectively. Our MCMC measured uncertainties appear to agree with the predictions in all cases, even at large values of $e$ and $b$ (see Section~\ref{sec:grazing} for a discussion of the effects of grazing transits on our approximations). At large $e$, the relative uncertainty in $\tau$ increases by more than an order of magnitude from the C08 prediction, highlighting the importance of accounting for finite integration in such cases. \red{Similarly, for small $P$, the relative uncertainty on $\tau$ increases significantly compared to previous predictions.}

Our analytic expressions for the covariances and variances clearly agree better with the results from the simulated \textit{Kepler} long cadence data than the finite cadence corrections from C08. The finite cadence corrections from C08 do not account for the averaging of the planet light curve over a finite integration time. In some cases where finite cadence corrections come to bear (e.g. in the variances of $T$ and $\tau$ as $R_p/R_\star$ gets smaller), finite integration time may actually improve the variances of $T$ and $\tau$ compared to the predictions of C08 Equation 26.  With finite integration time information on the ingress and egress is spread over any long exposure spanning the ingress and egress. In contrast, if the light curve is instantaneously sampled at the same cadence, the ingress and egress may be completely missed.

\section{\red{Survey planning}}
There are several space- and ground-based transit surveys on the horizon, including TESS, K2, and PLATO. Equations~\ref{eqn:cov1}, \ref{eqn:cov2}, \ref{eqn:cov3} and \ref{eqn:cov4} can be used to help choose an optimal integration time for photometric surveys for transiting planets when combined with models for the frame rate and photometric measurement uncertainty of the particular instrument.

In the Equations~\ref{eqn:cov1}, \ref{eqn:cov2}, \ref{eqn:cov3} and \ref{eqn:cov4}, the photometric precision $\sigma$ and integration time $\mathcal{I}$ are separate parameters. In practice, the uncertainty on a given photometric point will depend on the integration time chosen. For photon-noise, $\sigma/f_0 \propto \mathcal{I}^{-1/2}$. We have kept our equations explicitly in terms of $\sigma$ (instead of directly substituting in the assumption of photon-noise) so that they can be more flexibly applied to cases where additional white noise sources add to the photometric measurement uncertainty.

The integration time also affects the effective phase sampling of the light curve. For continuous photometric observations over a time baseline, $T_\mathrm{tot}$, the effective sampling of the phase-folded light curve can be up to

\begin{equation}
\Gamma_\mathrm{eff} = \frac{T_\mathrm{tot}}{P\left(\mathcal{I}+t_\mathrm{read}\right)}.
\label{eqn:gamma_eff}
\end{equation}

\noindent We denote by $t_\mathrm{read}$ the time needed for the photometer to read out; $\left(\mathcal{I}+t_\mathrm{read}\right)^{-1}$ is the CCD frame rate. The factor $T_\mathrm{tot} / P$ accounts for the number of transits detected over the span of the observations. This maximal effective sampling rate is achieved in the case where there is no clustering of photometric points at specific orbital phases.

\begin{figure*}
\begin{center}
\includegraphics[width=\linewidth]{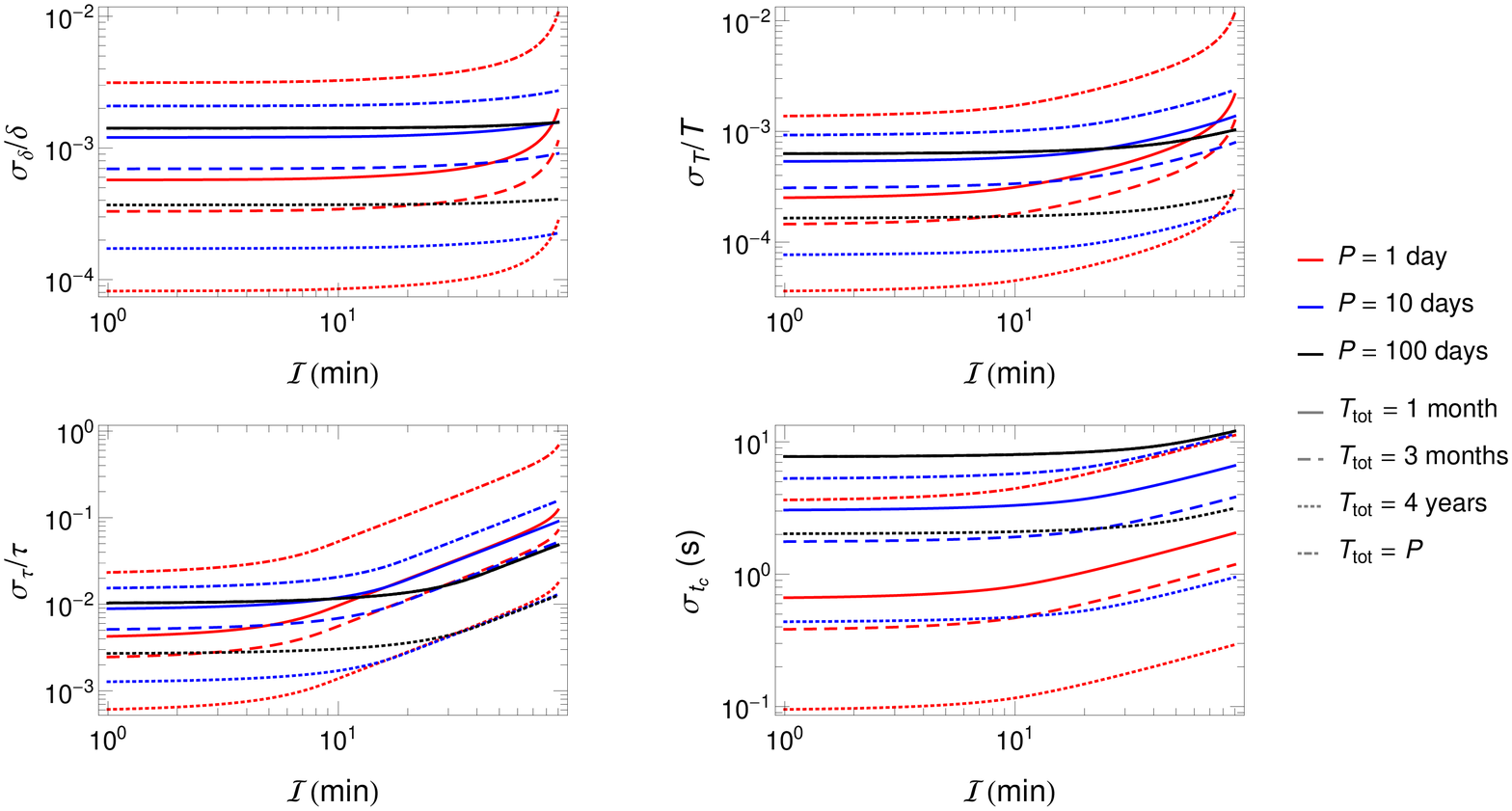}
\caption{\label{fig:survey_jupiter}
Uncertainties of trapezoidal transit parameters for representative observing cases as functions of integration time $\mathcal{I}$, as predicted by the binned trapezoidal light curve model, for Jupiter-size planets orbiting Sun twins. We show cases for orbits of 1 day (red curves), 10 days (blue curves), and 100 days (black curves) and total survey lengths of 1 month (TESS-like; solid curves), 3 months (dashed curves), 4 years ({\it Kepler}-like; dotted curves), \red{and one orbital period (dot-dashed curves)}. We assume nominal parameter values $b = 0.2$, $e = 0.16$, and $r = 0.1$, to remain consistent with the figures above; we scale the photometric uncertainty such that a \textit{Kepler} 30-minute exposure corresponds to $\sigma = 5 \times 10^{-5}$, and we assume photon-noise is the only noise source. $\Gamma_\mathrm{eff}$ is given by Equation \ref{eqn:gamma_eff} when $T_\mathrm{tot} > P$; otherwise, $\Gamma_\mathrm{eff} = 1/\mathcal{I}$. \red{The special case $T_\mathrm{tot} = P$ corresponds to the precision on the parameters derived from fitting a single transit light curve.}}
\end{center}
\end{figure*}

\begin{figure*}
\begin{center}
\includegraphics[width=\linewidth]{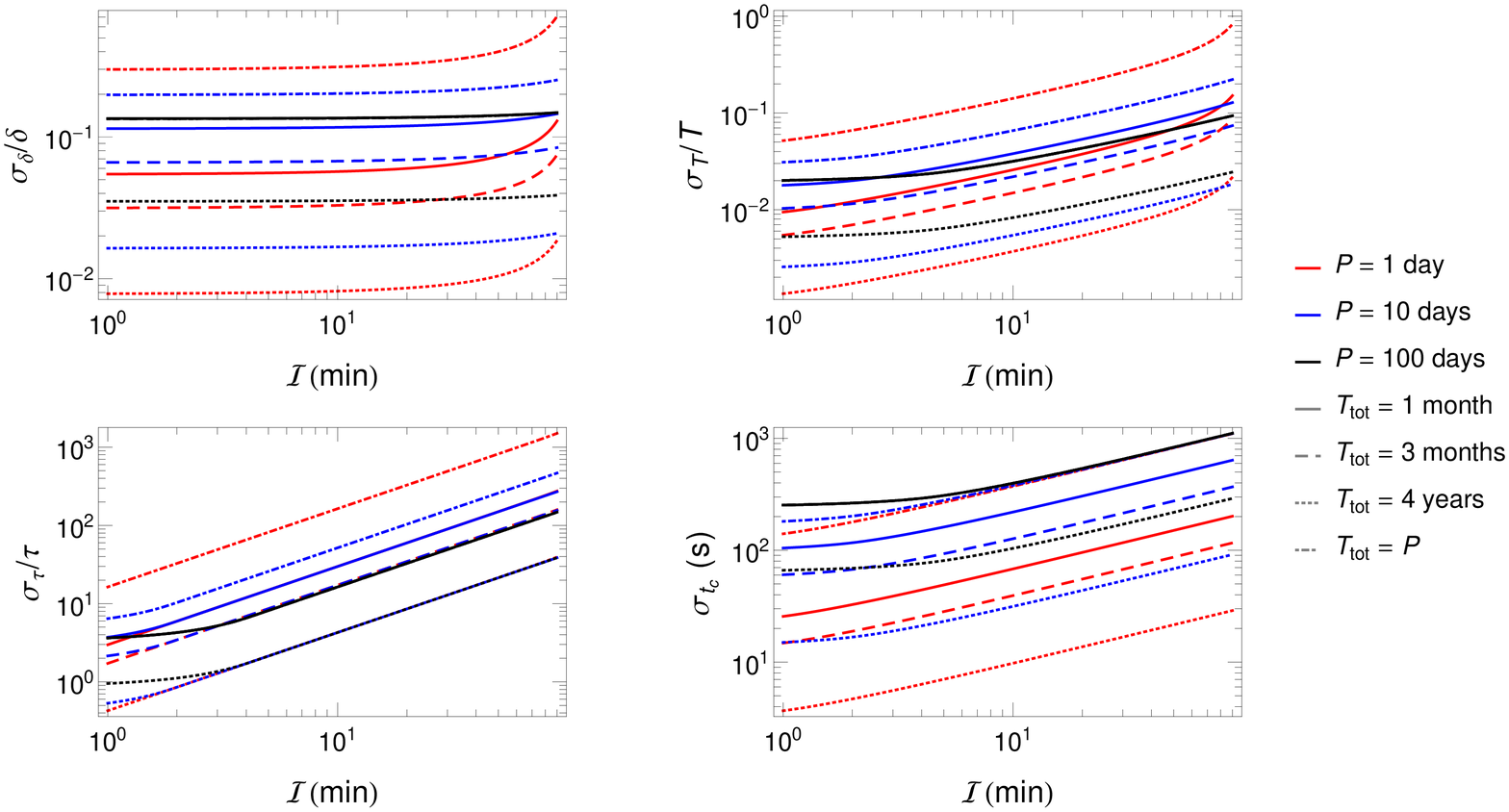}
\caption{\label{fig:survey_earth}
\red{Uncertainties of trapezoidal transit parameters for representative observing cases as functions of integration time $\mathcal{I}$, as predicted by the binned trapezoidal light curve model, for Earth-size planets orbiting Sun twins. We show cases for orbits of 1 day, 10 days, and 100 days and total survey lengths of 1 month, 3 months, 4 years, and one orbital period. We assume nominal parameter values $b = 0.2$, $e = 0.16$, and $r = 0.01$ and scale the photometric uncertainty such that a \textit{Kepler} 30-minute exposure corresponds to $\sigma = 5 \times 10^{-5}$, and we assume photon-noise is the only noise source. $\Gamma_\mathrm{eff}$ is given by Equation \ref{eqn:gamma_eff} when $T_\mathrm{tot} > P$; otherwise, $\Gamma_\mathrm{eff} = 1/\mathcal{I}$. The line styles and colors are identical to those in Figure~\ref{fig:survey_jupiter}.}}
\end{center}
\end{figure*}

In the case of a photon-noise limited survey with negligible read out time $\left(t_\mathrm{read}=0\right)$, the $\mathcal{I}$ dependence cancels in the $\sigma^2/\Gamma_\mathrm{eff}$ prefactor that scales all the covariance matrices. In these limits, the integration time dependence comes solely from the body of the covariance matrix elements in Equations~\ref{eqn:cov1}, \ref{eqn:cov2}, \ref{eqn:cov3} and \ref{eqn:cov4}. We plot in Figure~\ref{fig:survey_jupiter} how the uncertainties on the transit parameters predicted for a Jupiter transiting a Sun-twin in a photon-noise limited survey with negligible read-out time depend on $\mathcal{I}$, $T_\mathrm{tot}$, and $P$. We assumed a nominal photometric precision of $\sigma/f_0=5\times10^{-5}$ for a 30 minute exposure; choosing a different value for $\sigma$ would simply amount to rescaling the vertical axis on the figure. \red{In Figure~\ref{fig:survey_earth}, we make the same plot for a $1~R_\oplus$ planet orbiting a Sun-twin, assuming the same nominal orbit parameters.}

Lower integration times mean better precision, but after a certain point there is a plateau regime in which shorter integration times do not improve the relative precision of the transit parameters derived from the light curve (see Figures~\ref{fig:survey_jupiter} and \ref{fig:survey_earth}). In planning a transit survey, choosing an integration time near the ``knee'' would be optimal to minimize both the data rate and the relative uncertainties on the derived planet properties. The critical integration time depends on both the planet orbital period and $R_p/R_*$, but is not significantly affected by the survey duration, $T_\mathrm{tot}$. Planets with shorter $P$ and smaller $R_p/R_*$ have smaller critical integration times, and their characterization would benefit more greatly from short cadence observations.

The critical integration time delimiting the beginning of the plateau regime is different for different transit parameters of interest. The critical integration time for $\tau$ is the shortest. In planning a survey, one would want to consider the smallest critical integration time among the parameters of interest. Exposure times of 3, 10, and 30 minutes are optimal for sampling the transits of Jupiter-sized planets orbiting Sun twins on 1-day, 10-day, and 100-day orbits, respectively (Figure~\ref{fig:survey_jupiter}). In contrast, for Earth-sized planets with $R_p/R_*=10^{-2}$ the plateau in the relative uncertainty in $\tau$ occurs at $\mathcal{I}<1~\rm{minute}$ (Figure~\ref{fig:survey_earth}), regardless of the orbital period.

\section{Discussion}
\label{sec:dis}
In this section we revisit some of the approximations involved in deriving our covariance matrices, exploring their effects and quantifying the limitations they impose on the applicability of Equations \ref{eqn:cov1}, \ref{eqn:cov2}, \ref{eqn:cov3}, and \ref{eqn:cov4}.

\subsection{Effect of grazing transits}
\label{sec:grazing}
We have thus far limited our discussion to cases in which the integration time does not exceed the time between second and third contact, when the planet disk is contained completely within the disk of the star. Once the integration time exceeds $T-\tau$, the maximum apparent depth of the transit light curve starts to decrease as all exposures taken during totality are diluted by flux during ingress, egress, and/or out-of-transit. The apparent maximum depth of the transit light curve can be reduced by as much as $T/\mathcal{I}$; this maximum value corresponds to $\mathcal{I}>T+\tau$.

\begin{figure}
\begin{center}
\includegraphics[width=\columnwidth]{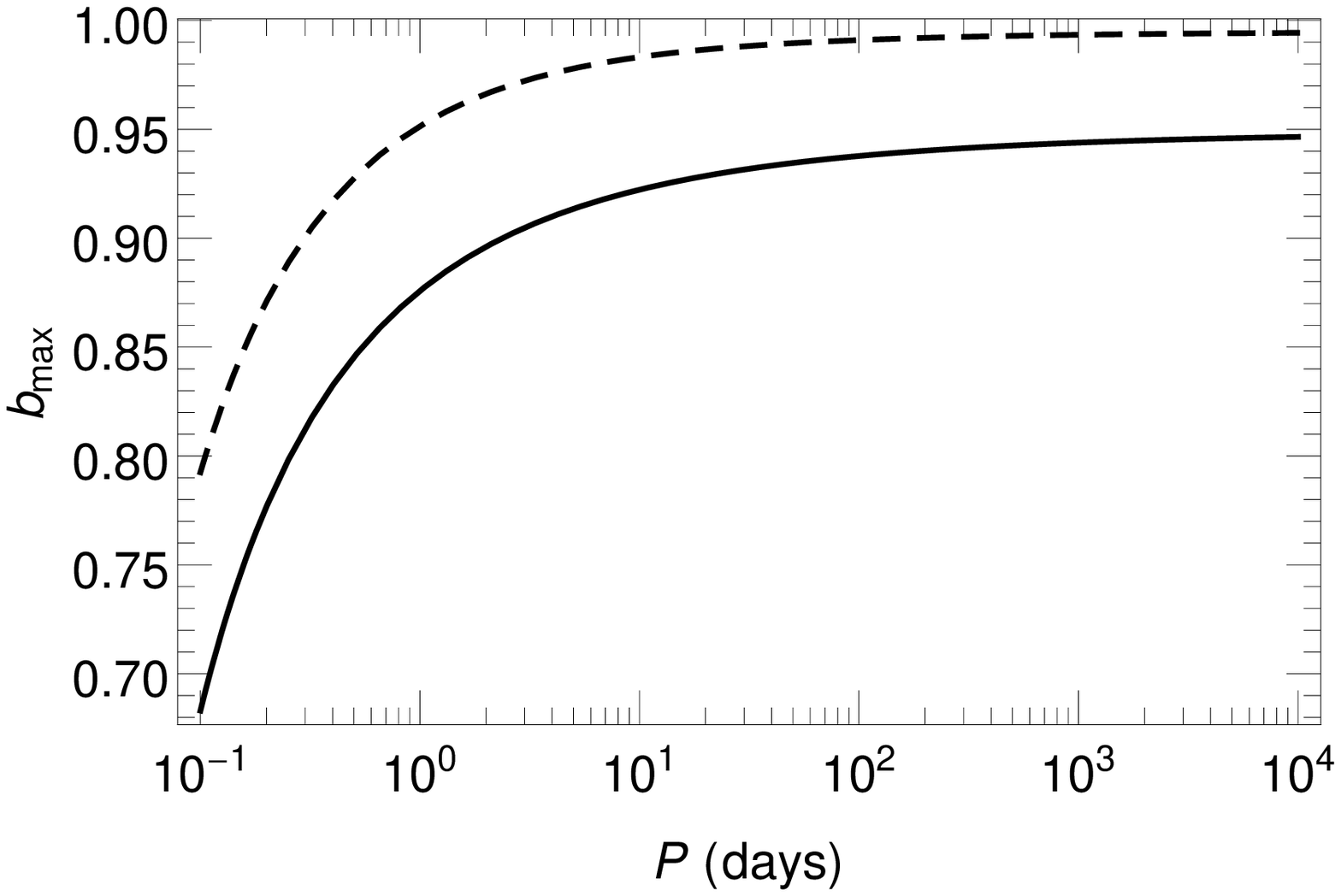}
\caption{\label{fig:bmax}
Maximum impact parameter, $b_{max}$, as a function of orbital period for Jupiter-sized ($r=0.1$, solid line) and Earth-sized ($r=0.01$, dashed) planets on circular orbits around a Sun-twin star with {\it Kepler} long-cadence sampling ($\mathcal{I}=30~\rm{minutes}$).}
\end{center}
\end{figure}

By focusing on cases with $\mathcal{I}<T-\tau$, we are effectively limiting the range of orbital inclinations (or impact parameters, $b$) considered. We are neglecting grazing transits, with impact parameters $b$ in excess of a maximum value, $b_{max}$, given by combining Equations~\ref{eq:T} and \ref{eq:tau},
\begin{equation}
b_{max}= \left[1-r\left(1+\frac{x^2}{2r}+\sqrt{\left(\frac{x^2}{2r}\right)\left(\frac{x^2}{2r}+2\right)}\right)\right]^{1/2},
\end{equation}

\noindent where $x=\frac{\mathcal{I}}{2\tau_0}$. In the case of long orbital periods ($x^2\ll 2r$) $b_{max}$ approaches $1-r$, the limiting value for the disk of the planet to be circumscribed by the stellar disk. For shorter periods ($\lesssim 10$ days) the constraints on $b$ for which our results apply can be significantly more restrictive (Figure~\ref{fig:bmax}). Limb-darkening could have an important effect on these grazing transits, and may lead to the breakdown of the C08 trapezoidal light curve approximations anyway, in the excluded regime of $b>b_{max}$.

Though we have not provided analytic equations for the covariance matrix in the case where $\mathcal{I}>T-\tau$, these can be readily derived following a similar approach as in Appendix~\ref{adx:fullFI}, below. There are in fact, three more regimes to be considered (in addition to cases 1 and 2 given in \red{Equation~\ref{eqn:lcbinned}}): $T-\tau<\mathcal{I}<T+\tau$ and $\mathcal{I}<\tau$ (case 3); $T-\tau<\mathcal{I}<T+\tau$ and $\mathcal{I}>\tau$ (case 4); and $\mathcal{I}>T+\tau$ (case 5).

\subsection{Effect of limb darkening}
\red{So far, we have neglected the effect of limb-darkening (following C08), and have considered a planet transiting a star with uniform surface brightness. To explore the effects of limb darkening, we generated synthetic transit data with a Python implementation of the \citet{EastmanEt2013ASP} EXOFAST \texttt{occultquad} routine, which generates a \citet{Mandel&Agol2002ApJ} quadratically limb-darkened light curve. We chose the limb darkening parameters for HAT-P-2 as our test case, obtaining the parameters with the \citet{EastmanEt2013ASP} limb darkening parameter applet\footnote{\url{http://astroutils.astronomy.ohio-state.edu/exofast/limbdark.shtml}}, which interpolates the \citet{ClaretEt2011AA} quadratic limb darkening tables.}

\red{We fit each synthetic data set with three different models (the trapezoidal model described in this paper, a Mandel \& Agol model with fixed limb darkening parameters, or a Mandel \& Agol model with limb darkening coefficients as free parameters). We used a procedure similar to that described in Section \ref{sec:MCMC} but with $1.5 \times 10^5$ samples. To minimize the covariances between model parameters, we parametrized all models in terms of $\delta$, $T$, $\tau$, and $t_c$. The limb darkened model also included the \citet{Kipping2013MNRAS} limb darkening parameters $q_1$ and $q_2$, which map directly to the quadratic coefficients $u_1$ and $u_2$, as free parameters. Since fitting eccentricity $e$ and argument of periastron $\omega$ is difficult and computationally intensive, we have restricted this test to $e = 0$ cases only.}

\begin{figure*}
\begin{center}
\includegraphics[width=\linewidth]{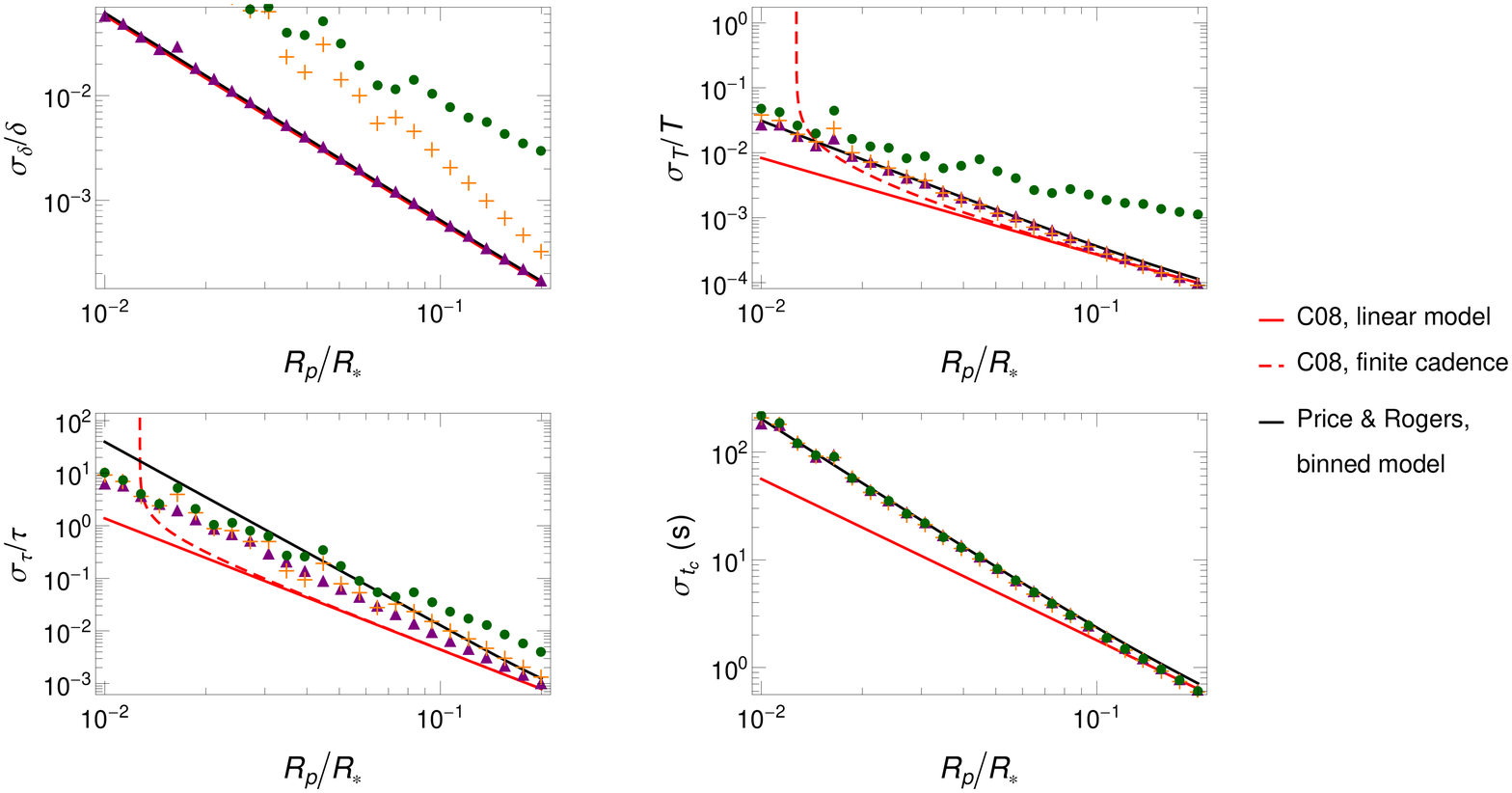}
\caption{\label{fig:varRpTrapzLD}
\red{Relative uncertainties of the trapezoidal transit parameters for a \citet{Mandel&Agol2002ApJ} quadratically limb darkened light curve, integrated with $\mathcal{I} = 30~\text{minutes}$, as a function of $R_p/R_\star$. The uncertainties were measured with an MCMC analysis, with the relative uncertainties being scaled by the true value of the parameter. We fit synthetic light curves with the trapezoidal model described in this paper (purple triangles), a Mandel \& Agol model with fixed limb darkening parameters (orange crosses), and a Mandel \& Agol model with limb darkening coefficients as free parameters (green points). Our predictions (solid black lines) are applicable for the trapezoidal model, though the uncertainty in $\tau$ is overpredicted. The uncertainty in $\delta$ increases when limb darkening is taken into account. The C08 prediction (solid red line) and finite cadence prediction (dashed red line) are shown for comparison. We let $R_\star = R_\odot$, $M_\star = M_\odot$, $e = 0$, $P = 9.55~\text{days}$, and $b = 0.2$. As before, the uncertainties are scaled by the true value of the parameter, so systematic offsets are not reflected in this plot.}}
\end{center}
\end{figure*}

\begin{figure*}
\begin{center}
\includegraphics[width=\linewidth]{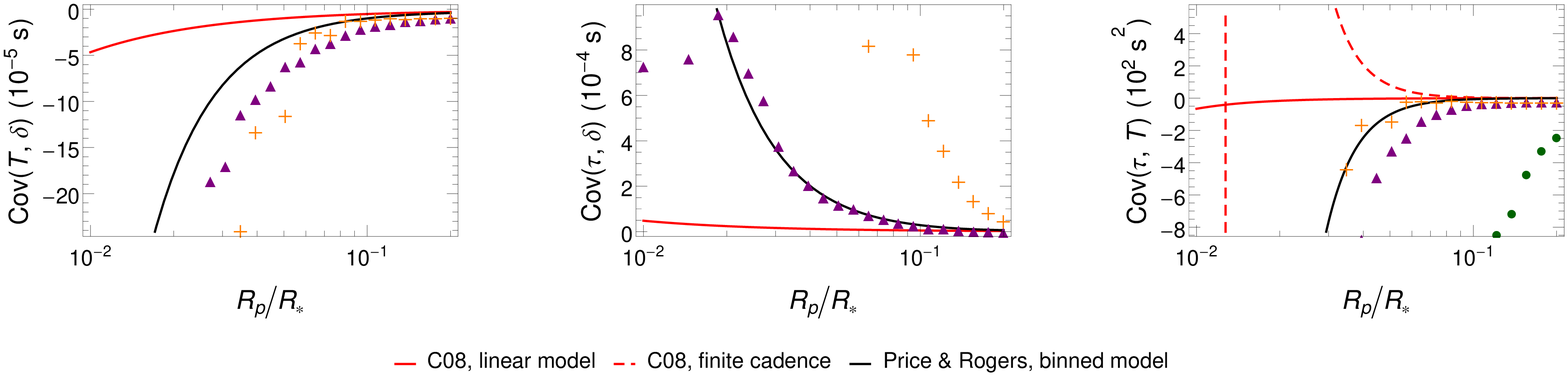}
\caption{\label{fig:covRpTrapzLD}
\red{Covariances in the trapezoidal approximation transit parameters derived from simulated \textit{Kepler} long-cadence data, as a function of $R_p/R_\star$, corresponding to the same scenarios presented in Figure~\ref{fig:varRpTrapzLD}. Fitting with a binned trapezoidal model yields values of $T$ that are significantly more correlated with $\delta$ and $\tau$ than we predict at smaller values of $R_p/R_*$. Our predictions are much less useful for predicting the parameter covariances of the Mandel \& Agol models.}}
\end{center}
\end{figure*}

\red{We show the results of this analysis in Figures \ref{fig:varRpTrapzLD} and \ref{fig:covRpTrapzLD}. In the case of the trapezoidal model, our Equations \ref{eqn:cov1}, \ref{eqn:cov2}, \ref{eqn:cov3}, and \ref{eqn:cov4} do well to predict the uncertainties on $\delta$, $T$, and $t_c$, but we overpredict the uncertainty in $\tau$. $T$ becomes more correlated with $\delta$ and with $\tau$ when limb darkening is taken into account. When the synthetic data are fit with a Mandel \& Agol model with fixed limb darkening, the uncertainty in the transit depth $\delta$ increases. Our predictions apply well to $T$ and $t_c$ in this case, while the uncertainty in $\tau$ is still overpredicted. Finally, when limb darkening coefficients are added as free parameters, the uncertainty in $\delta$ increases further, and we underpredict the uncertainty in $T$ significantly, as expected. To accurately predict the uncertainties in this case, $q_1$ and $q_2$ should be included as parameters in the Fisher information analysis. The uncertainty in $\tau$ is overpredicted in the small $R_p/R_\star$ regime and underpredicted in the large $R_p/R_\star$ regime. When a Mandel \& Agol model is used, we find that our equations are not as useful for predicting the parameter covariances.}

\subsection{Effect of finite phase sampling}
\label{sec:PhaseSampling}
Our analysis makes the approximation that the data is sampled at a uniform rate. The $\Gamma_\mathrm{eff}$ that we have defined for phase-folded data (Equation~\ref{eqn:gamma_eff}) is the maximum possible sampling rate for a continuous photometric time series. The approximation of a constant effective sampling rate for phase-folded data may break down if the planet orbital period is an integer (or rational number) multiple of the sampling cadence. In these scenarios, the photometric observations of different transits cluster at specific phases within the planet orbit and transit light curve, and the equal weighting of different phases in Equation~\ref{eqn:FisherElementIntegralFolded} is no longer fully valid. This will produce scatter about our analytic expressions for the covariance matrix in the idealized sampling scenario.

Another obstacle to applying our variance and covariance approximations arises if too few transits have been observed to sufficiently cover the full range of planet phases during transit. In these cases, the integral approximation of equation~\ref{eqn:FisherElementIntegral} breaks down and the finite sums (Equation~\ref{eqn:FisherElementSum}) must be evaluated numerically.

\subsection{Effect of non-Gaussian posteriors}
\label{sec:NonGaussian}
The posterior distribution of the trapezoidal parameters obtained from our MCMC fits to simulated {\it Kepler} long-cadence light curves are well approximated by Gaussians in high S/N scenarios. In cases of low S/N, however, the ``normally-distributed parameters'' assumption upon which the Fisher information matrix analysis relies can break down. The ingress/egress duration is physically constrained to be $\tau > 0$. When the uncertainty $\sigma_{\tau}$ becomes comparable to the magnitude of $\tau$ itself, the truncation at $\tau = 0$ induces non-Gaussian posteriors. This may account for some of the deviations at small $R_p/R_\star$ in Figures \ref{fig:varRpTrapz} and \ref{fig:covRpTrapz}. Solving numerically for the value of $R_p/R_\star$ where the value of $\tau$ becomes comparable to $\sigma_\tau$ for nominal values of the other parameters, we find that $\tau$ is equal to $3\sigma_\tau$ at $R_p/R_\star \approx 0.042$; it is equal to $2\sigma_\tau$ when $R_p/R_\star \approx 0.037$ and $1 \sigma_\tau$ when $R_p/R_\star \approx 0.031$. As $R_p/R_\star \rightarrow 0$, we expect the posterior to approach a Gaussian centered at $0$ and truncated at $0$; $\tau = 3\sigma_\tau$ is the approximate lower limit of $\tau$ down to which truncation should not be apparent in the posterior distribution. The numerical results seem to coincide well with the value of $R_p/R_\star$ where the MCMC results begin to deviate from the analytic predictions.

\subsection{Effect of other noise sources}
\label{sec:noise}
\red{Throughout this work, we have neglected the effects of correlated (red) noise on the light curve parameter uncertainties. For light curve data with significant red noise, we expect that our formulae will be less applicable, since we assume completely uncorrelated errors in the Fisher information analysis.}

\red{We have also neglected the effects of read noise $\sigma_\mathrm{read}$, noise that is intrinsic to the detector. Read noise adds in quadrature with photon-noise, thereby setting a minimum value of the overall uncertainty on each data point in the light curve. For shorter integration times $\mathcal{I}$ and fainter stars, read noise may dominate the overall noise $\sigma$. In the $\sigma \to \sigma_\mathrm{read}$ limit, $\sigma \sim \mathrm{constant}$, $\Gamma_\mathrm{eff} \sim 1 / \mathcal{I}$, and $f_0 \sim \mathcal{I}$, so all of the covariance elements we derive in Appendix~\ref{adx:fullFI} will be scaled by $\left( \sigma / f_0 \right)^2 / \Gamma_\mathrm{eff} \sim 1 / \mathcal{I}$. Thus, in the read-noise dominated regime, the uncertainties on transit parameters are expected to increase with shorter exposure times. In practice, the optimal integration time for a given target (assuming white noise) will be either the integration time at which read noise becomes the dominant noise source or the critical integration time assuming photon noise (the ``knee'' in Figures 8 and 9), whichever is longer.}

\section{Conclusions}
\label{sec:con}
\citet{Kipping2010MNRAS} highlighted the necessity of fitting a binned light curve model to binned light curve data. We have updated the \citet{CarterEt2008ApJ} analytic expressions for the variances and covariances of parameters derived from fitting transit light curve data, to take finite integration time into account.

With finite integration time, the uncertainties on the transit parameters are strictly greater than what one could extract from an instantaneously sampled light curve. The magnitude of the correlations among transit ingress/egress duration, transit duration and transit depth all increase, while the mid-transit time (relevant for measuring TTVs) remains uncorrelated. For example, for a Hot Jupiter or close-in Earth-size planet on a three day orbit the variances on $\delta$, $t_c$, $\tau$, and $T$ are $1.2$, $2.5$, $24$, and $2.8$ times larger for 30-minute long-cadence data as compared to 1 minute short cadence data, \red{assuming the nominal orbit parameters we have used throughout this work}; the covariances can increase by as much as a factor of $30$. In contrast, for a transiting Earth-twin on a 1-year orbit, the variances themselves are larger in magnitude, but they do not change greatly with integration time.

We provide Python and \textit{Mathematica} code for computing the predicted variances and covariances that could be measured using the binned light curve model. Phasing, red noise, non-Gaussianities, and other effects can affect the actual uncertainties obtained from a full analysis. Our analytic expressions are still helpful for target selection, observation planning, and rule of thumb intuition. Today, finite integration time is relevant for \textit{Kepler} long-cadence light curves, and will remain important in the future analysis of data from K2 and from TESS full frame images.

\section{Acknowledgements}
We would like to thank John Johnson of the Harvard-Smithsonian Center for Astrophysics for his valuable input on this project and for establishing the Johnson Exolab as an environment where undergraduates and postdoctoral scholars can work together on projects like this one. We would also like to thank the referee for providing a very helpful and constructive review of this work. EMP acknowledges funding provided by Mr. and Mrs. Carl Larson for her 2013 Carolyn Ash SURF Fellowship. LAR acknowledges support provided by NASA through Hubble Fellowship grant \#HF-51313.01 awarded by the Space Telescope Science Institute, which is operated by the Association of Universities for Research in Astronomy, Inc., for NASA, under contract NAS 5-26555. The LevelScheme \citep{Caprio2005CPC} scientific figure preparation system for \textit{Mathematica} was used in the preparation of this paper.

\appendix

\section{Full derivation of Fisher information and covariance matrices}
\label{adx:fullFI}
For $N$ data points $\{ y \}$ and model points $\{ y_{\rm mod} \}$ which depend on a set of parameters $\{p\}$, and under the assumption of uncorrelated errors of constant absolute magnitude $\sigma$, we begin with the likelihood function \citep[see e.g.,][]{Gould2003}

\begin{equation}
\mathcal{L} = \frac{1}{\sigma \sqrt{2\pi}} \exp{\left[ -\frac{1}{2} \sigma^{-2} \sum\limits_{k=1}^{N} \left(y_k - y_{k,\text{mod}} \right)^2 \right]}.
\label{eqn:likelihood}
\end{equation}

\noindent The Fisher information matrix $\mathbf{B}$ is defined by \citep[see e.g.,][]{Vallisneri2008},

\begin{align}
B_{ij} &= \langle \left( \frac{\partial}{\partial p_i} \log{\mathcal{L}} \right) \left( \frac{\partial}{\partial p_j} \log{\mathcal{L}} \right) \rangle \nonumber \\
&= \langle \left( \sigma^{-4} \sum\limits_{k=1}^N \sum\limits_{l=1}^N \left( y_k - y_{k,\text{mod}} \right) \left( y_l - y_{l,\text{mod}} \right) \frac{\partial y_{k,\text{mod}}}{\partial p_i} \frac{\partial y_{l,\text{mod}}}{\partial p_j} \right) \rangle \nonumber \\
&= \sigma^{-2} \sum\limits_{k=1}^N \left( \frac{\partial y_{k,\text{mod}}}{\partial p_i} \right) \left( \frac{\partial y_{k,\text{mod}}}{\partial p_j} \right)
\end{align}

\noindent \red{where $\langle x \rangle$ denotes the expected value of $x$.} For our application to transit light curves, the model function is $y_{mod} = F_{lb}$ and $p = \{t_c, \tau, T, \delta, f_0\}$, so

\begin{equation}
B_{ij} = \sigma^{-2} \sum\limits_{k=1}^{N} \left[ \frac{\partial}{\partial p_i} F_{lb}(t_k;\,\{p\}) \right] \left[ \frac{\partial}{\partial p_j} F_{lb}(t_k;\,\{p\}) \right].
\label{eqn:FisherElementSum}
\end{equation}

\noindent \red{Tables \ref{tab:DerivativesF1} and \ref{tab:DerivativesF2} give partial derivatives for the five regions of the binned light curve model for $F_{lb1}$ and $F_{lb2}$, where $F_{lb1}$ and $F_{lb2}$ are equivalent to $F_{lb}$ limited to the $\tau > \mathcal{I}$ and $\tau < \mathcal{I}$ regimes, respectively.}

We assume that the data points are sampled uniformly with a uniform sampling rate $\Gamma$, beginning at time point $t_0$ and for a total duration $T_\mathrm{tot}$. Like C08, we approximate the finite sums of Equation~\ref{eqn:FisherElementSum} by an integral over time, assuming that $\Gamma$ is large enough to sufficiently sample the transit light curve:

\begin{equation}
B_{ij} = \frac{\Gamma}{\sigma^2} \int\limits_{t_0}^{t_0+T_\mathrm{tot}} \left[ \frac{\partial}{\partial p_i} F(t;\,\{p\}) \right] \left[ \frac{\partial}{\partial p_j} F(t;\,\{p\}) \right] dt.
\label{eqn:FisherElementIntegral}
\end{equation}

\noindent Substituting $\Gamma_\mathrm{eff}$ for $\Gamma$ and $P$ for $T_\mathrm{tot}$, as described in Section~\ref{sec:FI} for phase-folded data, this equation becomes

\begin{equation}
B_{ij} = \frac{\Gamma_\mathrm{eff}}{\sigma^2} \int\limits_{t_0}^{t_0+P} \left[ \frac{\partial}{\partial p_i} F(t;\,\{p\}) \right] \left[ \frac{\partial}{\partial p_j} F(t;\,\{p\}) \right] dt.
\label{eqn:FisherElementIntegralFolded}
\end{equation}

Evaluating Equation \ref{eqn:FisherElementIntegral} with the partial derivatives given in Tables \ref{tab:DerivativesF1} and \ref{tab:DerivativesF2} yields the Fisher matrices in Equations \ref{eqn:FisherMatrix1} and \ref{eqn:FisherMatrix2}. In the $\tau > \mathcal{I}$ case, we find the Fisher information matrix to be

\begin{equation}
B_{lb1} = \frac{\Gamma}{\sigma^2} \left(
\begin{array}{ccccc}
 -\frac{2 \delta ^2 (\mathcal{I}-3 \tau )}{3 \tau ^2} & 0 & 0 & 0 & 0 \\
 0 & \frac{\delta ^2 \left(\mathcal{I}^3-5 \tau ^2 \mathcal{I}+5 \tau ^3\right)}{30 \tau ^4} & 0 & -\frac{\delta  \left(2 \mathcal{I}^3-5 \tau  \mathcal{I}^2+10 \tau ^3\right)}{60 \tau ^3} & 0 \\
 0 & 0 & -\frac{\delta ^2 (\mathcal{I}-3 \tau )}{6 \tau ^2} & \frac{\delta }{2} & -\delta  \\
 0 & -\frac{\delta  \left(2 \mathcal{I}^3-5 \tau  \mathcal{I}^2+10 \tau ^3\right)}{60 \tau ^3} & \frac{\delta }{2} & T+\frac{\mathcal{I}^3-5 \tau  \mathcal{I}^2-10 \tau ^3}{30 \tau ^2} & -T \\
 0 & 0 & -\delta  & -T & T_\mathrm{tot} \\
\end{array}
\right).
\label{eqn:FisherMatrix1}
\end{equation}

\noindent In the $\tau < \mathcal{I}$ case, we find

\begin{equation}
B_{lb2} = \frac{\Gamma}{\sigma^2} \left(
\begin{array}{ccccc}
 \frac{2 \delta ^2 (3 \mathcal{I}-\tau )}{3 \mathcal{I}^2} & 0 & 0 & 0 & 0 \\
 0 & \frac{\delta ^2 \tau }{30 \mathcal{I}^2} & 0 & \frac{\delta  \tau  (3 \tau -10 \mathcal{I})}{60 \mathcal{I}^2} & 0 \\
 0 & 0 & \frac{\delta ^2 (3 \mathcal{I}-\tau )}{6 \mathcal{I}^2} & \frac{\delta }{2} & -\delta  \\
 0 & \frac{\delta  \tau  (3 \tau -10 \mathcal{I})}{60 \mathcal{I}^2} & \frac{\delta }{2} & T+\frac{-10 \mathcal{I}^3-5 \tau ^2 \mathcal{I}+\tau ^3}{30 \mathcal{I}^2} & -T \\
 0 & 0 & -\delta  & -T & T_\mathrm{tot} \\
\end{array}
\right).
\label{eqn:FisherMatrix2}
\end{equation}

\begin{table*}
\caption{Variables defined for the simplification of the trapezoidal parameters covariance matrix of
Equation~\ref{eqn:cov1}.}
\begin{center}
\bgroup
\begin{tabular}{ll}
\hline\hline
Symbol & Expression \\
\hline
$a_1$ & $\left( 10 \tau ^3+2 \mathcal{I}^3-5 \tau  \mathcal{I}^2 \right) / \tau^3$ \\
$a_2$ & $\left( 5 \tau ^3+\mathcal{I}^3-5 \tau ^2 \mathcal{I} \right) / \tau^3$ \\
$a_3$ & $\left( 9 \mathcal{I}^5 T_\mathrm{tot}-40 \tau ^3 \mathcal{I}^2 T_\mathrm{tot}+120 \tau ^4 \mathcal{I} (3 T_\mathrm{tot}-2 \tau ) \right) / \tau^6$ \\
$a_4$ & $\left( a_3  \tau ^5+\mathcal{I}^4 (54 \tau -35 T_\mathrm{tot})-12 \tau  \mathcal{I}^3 (4 \tau +T_\mathrm{tot})+360 \tau ^4 (\tau -T_\mathrm{tot}) \right) / \tau^5$ \\
$a_5$ & $\left( a_2  \left(24 T^2 (\mathcal{I}-3 \tau )-24 T T_\mathrm{tot} (\mathcal{I}-3 \tau )\right)+\tau ^3 a_4 \right) / \tau^3$ \\
$a_6$ & $\left( 3 \tau ^2+T (\mathcal{I}-3 \tau ) \right) / \tau^2$ \\
$a_7$ & $\left( -60 \tau ^4+12 a_2  \tau ^3 T-9 \mathcal{I}^4+8 \tau  \mathcal{I}^3+40 \tau ^3 \mathcal{I} \right) / \tau^4$ \\
$a_8$ & $\left( 2 T-T_\mathrm{tot} \right) / \tau$ \\
$a_9$ & $\left( -3 \tau ^2 \mathcal{I} \left(-10 T^2+10 T T_\mathrm{tot}+\mathcal{I} (2 \mathcal{I}+5 T_\mathrm{tot})\right)-\mathcal{I}^4 T_\mathrm{tot}+8 \tau  \mathcal{I}^3 T_\mathrm{tot} \right) / \tau^5$ \\
$a_{10}$ & $\left( a_9  \tau ^2+60 \tau ^2+10 \left(-9 T^2+9 T T_\mathrm{tot}+\mathcal{I} (3 \mathcal{I}+T_\mathrm{tot})\right)-75 \tau  T_\mathrm{tot} \right) / \tau^2$ \\
$a_{11}$ & $\left( \mathcal{I} T_\mathrm{tot}-3 \tau  (T_\mathrm{tot}-2 \tau ) \right) / \tau^2$ \\
$a_{12}$ & $\left( -360 \tau ^5-24 a_2  \tau ^3 T (\mathcal{I}-3 \tau )+9 \mathcal{I}^5-35 \tau  \mathcal{I}^4-12 \tau ^2 \mathcal{I}^3-40 \tau ^3 \mathcal{I}^2+360 \tau ^4 \mathcal{I} \right) / \tau^5$ \\
$a_{13}$ & $\left( -3 \mathcal{I}^3 \left(8 T^2-8 T T_\mathrm{tot}+3 \mathcal{I} T_\mathrm{tot} \right)+120 \tau ^2 T \mathcal{I} (T-T_\mathrm{tot})+8 \tau  \mathcal{I}^3 T_\mathrm{tot} \right) / \tau^5$ \\
$a_{14}$ & $\left( a_{13} \tau ^2+40 \left(-3 T^2+3 T T_\mathrm{tot}+\mathcal{I} T_\mathrm{tot}\right)-60 \tau  T_\mathrm{tot} \right) / \tau^2$ \\
$a_{15}$ & $\left( 2 \mathcal{I}-6 \tau \right) / \tau$ \\
\hline
\end{tabular}
\egroup
\end{center}
\label{tab:cov1vars}
\end{table*}

\noindent The full covariance matrix for each model is found by taking the matrix inverse of the Fisher matrix. We define the variables in Table~\ref{tab:cov1vars} to simplify the covariance matrix in the $\tau > \mathcal{I}$ case, given in Equation~\ref{eqn:cov1}.

\begin{equation}
\text{Cov}\left( \{t_c, \tau, T, \delta, f_0\}, \{t_c, \tau, T, \delta, f_0\};~\tau > \mathcal{I} \right) =
\frac{\sigma ^2}{\Gamma } \left(
\begin{array}{ccccc}
 -\frac{3 \tau }{\delta ^2 a_{15}} & 0 & 0 & 0 & 0 \\
 0 & \frac{24 \tau  a_{10}}{\delta ^2 a_5} & \frac{36 a_8 \tau  a_1}{\delta ^2 a_5} & -\frac{12 a_{11} a_1}{\delta  a_5} & -\frac{12 a_6 a_1}{\delta  a_5} \\
 0 & \frac{36 a_8 \tau  a_1}{\delta ^2 a_5} & \frac{6 \tau  a_{14}}{\delta ^2 a_5} & \frac{72 a_8 a_2}{\delta  a_5} & \frac{6 a_7}{\delta  a_5} \\
 0 & -\frac{12 a_{11} a_1}{\delta  a_5} & \frac{72 a_8 a_2}{\delta  a_5} & -\frac{24 a_{11} a_2}{\tau  a_5} & -\frac{24 a_6 a_2}{\tau  a_5} \\
 0 & -\frac{12 a_6 a_1}{\delta  a_5} & \frac{6 a_7}{\delta  a_5} & -\frac{24 a_6 a_2}{\tau  a_5} & \frac{a_{12}}{\tau  a_5} \\
\end{array}
\right)
\label{eqn:cov1}
\end{equation}

\begin{table*}
\caption{Variables defined for the simplification of the trapezoidal parameters covariance matrix of
Equation~\ref{eqn:cov2}.}
\begin{center}
\bgroup
\begin{tabular}{ll}
\hline\hline
Symbol & Expression \\
\hline
$b_1$ & $\left( 6 \mathcal{I}^2-3 \mathcal{I} T_\mathrm{tot}+\tau  T_\mathrm{tot} \right) / \mathcal{I}^2$ \\
$b_2$ & $\left( \tau  T+3 \mathcal{I} (\mathcal{I}-T) \right) / \mathcal{I}^2$ \\
$b_3$ & $\left( \tau ^3-12 T \mathcal{I}^2+8 \mathcal{I}^3+20 \tau  \mathcal{I}^2-8 \tau ^2 \mathcal{I} \right) / \mathcal{I}^3$ \\
$b_4$ & $\left( 6 T^2-6 T T_\mathrm{tot}+\mathcal{I} (5 T_\mathrm{tot}-4 \mathcal{I}) \right) / \mathcal{I}^2$ \\
$b_5$ & $\left( 10 \mathcal{I} - 3 \tau \right) / \mathcal{I}$ \\
$b_6$ & $\left( 12 b_4 \mathcal{I}^3+4 \tau  \left(-6 T^2+6 T T_\mathrm{tot}+\mathcal{I} (13 T_\mathrm{tot}-30 \mathcal{I})\right) \right) / \mathcal{I}^3$ \\
$b_7$ & $\left( b_6 \mathcal{I}^5+4 \tau ^2 \mathcal{I}^2 (12 \mathcal{I}-11 T_\mathrm{tot})+\tau ^3 \mathcal{I} (11 T_\mathrm{tot}-6 \mathcal{I})-\tau ^4 T_\mathrm{tot} \right) / \mathcal{I}^5$ \\
$b_8$ & $\left( 3 T^2-3 T T_\mathrm{tot}+\mathcal{I} T_\mathrm{tot} \right) / \mathcal{I}^2$ \\
$b_9$ & $\left( 8 b_8 \mathcal{I}^4+20 \tau  \mathcal{I}^2 T_\mathrm{tot}-8 \tau ^2 \mathcal{I} T_\mathrm{tot}+\tau ^3 T_\mathrm{tot} \right) / \mathcal{I}^4$ \\
$b_{10}$ & $\left( -\tau ^4+24 T \mathcal{I}^2 (\tau -3 \mathcal{I})+60 \mathcal{I}^4+52 \tau  \mathcal{I}^3-44 \tau ^2 \mathcal{I}^2+11 \tau ^3 \mathcal{I} \right) / \mathcal{I}^4$ \\
$b_{11}$ & $\left( -15 b_4 \mathcal{I}^3+10 b_8 \tau  \mathcal{I}^2+15 \tau ^2 (2 \mathcal{I}-T_\mathrm{tot}) \right) / \mathcal{I}^3$ \\
$b_{12}$ & $\left( b_{11} \mathcal{I}^5+2 \tau ^3 \mathcal{I} (4 T_\mathrm{tot}-3 \mathcal{I})-\tau ^4 T_\mathrm{tot} \right) / \mathcal{I}^5$ \\
$b_{13}$ & $\left( T_\mathrm{tot}-2 T \right) / \mathcal{I}$ \\
$b_{14}$ & $\left( 6 \mathcal{I}-2 \tau \right) / \mathcal{I}$ \\
\hline
\end{tabular}
\egroup
\end{center}
\label{tab:cov2vars}
\end{table*}

\noindent Similarly, we define the variables in Table~\ref{tab:cov2vars} to express the covariance matrix in the $\tau < \mathcal{I}$ case, given in Equation~\ref{eqn:cov2}.

\begin{equation}
\text{Cov} \left(\{t_c, \tau, T, \delta, f_0\}, \{t_c, \tau, T, \delta, f_0\};~\tau < \mathcal{I} \right) =
\frac{\sigma ^2}{\Gamma} \left(
\begin{array}{ccccc}
 \frac{3 \mathcal{I}}{\delta ^2 b_{14}} & 0 & 0 & 0 & 0 \\
 0 & -\frac{24 \mathcal{I}^2 b_{12}}{\delta ^2 \tau  b_7} & \frac{36 \mathcal{I} b_{13} b_5}{\delta ^2 b_7} & \frac{12 b_5 b_1}{\delta  b_7} & \frac{12 b_5 b_2}{\delta  b_7} \\
 0 & \frac{36 \mathcal{I} b_{13} b_5}{\delta ^2 b_7} & \frac{6 \mathcal{I} b_9}{\delta ^2 b_7} & \frac{72 b_{13}}{\delta  b_7} & \frac{6 b_3}{\delta  b_7} \\
 0 & \frac{12 b_5 b_1}{\delta  b_7} & \frac{72 b_{13}}{\delta  b_7} & \frac{24 b_1}{\mathcal{I} b_7} & \frac{24 b_2}{\mathcal{I} b_7} \\
 0 & \frac{12 b_5 b_2}{\delta  b_7} & \frac{6 b_3}{\delta  b_7} & \frac{24 b_2}{\mathcal{I} b_7} & \frac{b_{10}}{\mathcal{I} b_7} \\
\end{array}
\right)
\label{eqn:cov2}
\end{equation}

\noindent \red{Equation~\ref{eqn:cov1}} reduces to the C08 covariance matrix (their Equation 20) in the limit that $\mathcal{I}\to0$.

Following C08, we transform the covariance matrices to a more physical parameter space, parameterized by the variables $t_c$, $b^2$, $\tau_0^2$, $r$, and $f_0$, given by the inverse mapping

\begin{align}
r &= \left( \frac{\delta}{f_0} \right)^{1/2} \label{eqn:r} \\
b^2 &= 1 - \frac{r T}{\tau} \label{eqn:bsq} \\
\tau_0^2 &= \frac{T \tau}{4 r} \label{eqn:tau0sq}.
\end{align}

\noindent The covariance matrix of the physical parameters is then found by the transformation

\begin{equation}
\text{Cov}'(...) = \mathbf{J}^T \text{Cov}(...) \mathbf{J}
\end{equation}

\noindent with $\mathbf{J}$ the Jacobian matrix

\begin{equation}
\mathbf{J} = \frac{\partial (t_c, b^2, \tau_0^2, r, f_0)}{\partial (t_c, \tau, T, \delta, f_0)} = \left(
\begin{array}{ccccc}
 1 & 0 & 0 & 0 & 0 \\
 0 & \frac{T r}{\tau ^2} & \frac{T}{4 r} & 0 & 0 \\
 0 & -\frac{r}{\tau } & \frac{\tau }{4 r} & 0 & 0 \\
 0 & -\frac{T}{2 f_0 r \tau } & -\frac{T \tau }{8 f_0 r^3} & \frac{1}{2 f_0 r} & 0 \\
 0 & \frac{T r}{2 f_0 \tau } & \frac{T \tau }{8 f_0 r} & -\frac{r}{2 f_0} & 1 \\
\end{array}
\right).
\end{equation}

\begin{table*}
\caption{Variables defined for the simplification of the physical parameters covariance matrix of
Equation~\ref{eqn:cov3}.}
\begin{center}
\bgroup
\begin{tabular}{ll}
\hline\hline
Symbol & Expression \\
\hline
$A_1$ & $\left(T^2 \left(24 f_0^2 (2 a_1 a_{11}+4 a_{10}-a_{11} a_2)+48 a_6 \delta  f_0 (a_2-a_1)+a_{12} \delta ^2\right) \right) / \left( f_0^2 \tau ^2 \right)$ \\
$A_2$ & $\left( 24 f_0^2 (a_{11} T (a_1-a_2)+6 a_2 a_8 \tau )-12 \delta  f_0 (2 a_6 T (a_1-2 a_2)+a_7 \tau )+a_{12} \delta ^2 T\right) / \left( f_0^2 \tau \right)$ \\
$A_3$ & $\left( 24 f_0^2 (a_{11} T (a_1-a_2)-6 a_2 a_8 \tau )+12 \delta  f_0 (a_7 \tau -2 a_6 T (a_1-2 a_2))+a_{12} \delta ^2 T \right) / \left( f_0^2 \tau \right)$ \\
$A_4$ & $\left( a_{12} \delta  T-12 f_0 (2 a_6 T (a_1-a_2)+a_7 \tau )\right) / \left( f_0 \tau \right)$ \\
$A_5$ & $\left( 12 f_0 (2 a_6 T (a_2-a_1)+a_7 \tau )+a_{12} \delta  T \right) / \left( f_0 \tau \right)$ \\
$A_6$ & $\left( 288 a_8 f_0^2 \tau  T (a_2-a_1)-24 a_7 \delta  f_0 \tau  T \right) / \left( f_0^2 \tau ^2 \right)$ \\
$A_7$ & $24 a_{14}$ \\
$A_8$ & $\left( a_{12} \delta +24 a_2 a_6 f_0 \right) / f_0$ \\
$A_9$ & $\left( a_{12} \delta ^2-24 a_2 f_0 (a_{11} f_0-2 a_6 \delta ) \right) / f_0^2$ \\
\hline
\end{tabular}
\egroup
\end{center}
\label{tab:cov3vars}
\end{table*}

\noindent As before, we define several variables so that we can write the covariance matrix compactly. For $\tau > \mathcal{I}$, they are given in Table~\ref{tab:cov3vars}. With these definitions, the transformed covariance matrix in the $\tau > \mathcal{I}$ case becomes

\begin{equation}
\text{Cov}(\{t_c,b^2,\tau_0^2,r,f_0\},\{t_c,b^2,\tau_0^2,r,f_0\};~\tau > \mathcal{I}) =
\frac{\sigma ^2}{\Gamma} \left(
\begin{array}{ccccc}
 -\frac{3 \tau }{f_0^2 r^4 a_{15}} & 0 & 0 & 0 & 0 \\
 0 & \frac{A_1+A_6+A_7}{4 f_0^2 r^2 a_5 \tau } & \frac{\tau  (A_1-A_7)}{16 f_0^2 r^4 a_5} & -\frac{A_2}{4 f_0^2 r^2 a_5 \tau } & \frac{A_4}{2 f_0 r a_5 \tau } \\
 0 & \frac{\tau  (A_1-A_7)}{16 f_0^2 r^4 a_5} & \frac{\tau ^3 (A_1-A_6+A_7)}{64 f_0^2 r^6 a_5} & -\frac{\tau  A_3}{16 f_0^2 r^4 a_5} & \frac{\tau  A_5}{8 f_0 r^3 a_5} \\
 0 & -\frac{A_2}{4 f_0^2 r^2 a_5 \tau } & -\frac{\tau  A_3}{16 f_0^2 r^4 a_5} & \frac{A_9}{4 f_0^2 r^2 a_5 \tau } & -\frac{A_8}{2 f_0 r a_5 \tau } \\
 0 & \frac{A_4}{2 f_0 r a_5 \tau } & \frac{\tau  A_5}{8 f_0 r^3 a_5} & -\frac{A_8}{2 f_0 r a_5 \tau } & \frac{a_{12}}{a_5 \tau } \\
\end{array}
\right).
\label{eqn:cov3}
\end{equation}

\begin{table*}
\caption{Variables defined for the simplification of the physical parameters covariance matrix of
Equation~\ref{eqn:cov4}.}
\begin{center}
\bgroup
\begin{tabular}{ll}
\hline\hline
Symbol & Expression \\
\hline
$B_1$ & $\left( -24 b_1 f_0^2 \tau ^2 T^2 (2 b_5 \mathcal{I}-\tau )+b_{10} \delta ^2 \tau ^3 T^2-96 b_{12} f_0^2 T^2 \mathcal{I}^3+48 b_2 \delta  f_0 \tau ^2 T^2 (b_5 \mathcal{I}-\tau ) \right) / \left( f_0^2 \tau ^5 \right)$ \\
$B_2$ & $\left( b_{10} \delta  \tau  T+24 b_2 f_0 T (b_5 \mathcal{I}-\tau )-12 b_3 f_0 \tau  \mathcal{I} \right) / \left( f_0 \tau ^2 \right)$ \\
$B_3$ & $\left( b_{10} \delta  \tau  T+24 b_2 f_0 T (b_5 \mathcal{I}-\tau )+12 b_3 f_0 \tau  \mathcal{I} \right) / \left( f_0 \tau ^2 \right)$ \\
$B_4$ & $\left( 24 b_2 f_0-b_{10} \delta \right) / f_0$ \\
$B_5$ & $\left( 24 b_9 \mathcal{I}^2 \right) / \tau ^2$ \\
$B_6$ & $\left( 288 b_{13} f_0^2 T \mathcal{I} (\tau -b_5 \mathcal{I})-24 b_3 \delta  f_0 \tau  T \mathcal{I} \right) / \left( f_0^2 \tau ^3 \right)$ \\
$B_7$ & $\left( 24 b_5 f_0 T \mathcal{I} (b_1 f_0-b_2 \delta )-\tau  \left(24 f_0^2 (b_1 T+6 b_{13} \mathcal{I})+b_{10} \delta ^2 T-12 \delta  f_0 (4 b_2 T+b_3 \mathcal{I})\right) \right) / \left( f_0^2 \tau ^2 \right)$ \\
$B_8$ & $\left( 24 b_5 f_0 T \mathcal{I} (b_1 f_0-b_2 \delta )-\tau  \left(24 f_0^2 (b_1 T-6 b_{13} \mathcal{I})+b_{10} \delta ^2 T+12 \delta  f_0 (b_3 \mathcal{I}-4 b_2 T)\right) \right) / \left( f_0^2 \tau ^2 \right)$ \\
$B_9$ & $\left( 24 b_1 f_0^2+b_{10} \delta ^2-48 b_2 \delta  f_0 \right) / f_0^2$ \\
\hline
\end{tabular}
\egroup
\end{center}
\label{tab:cov4vars}
\end{table*}

\noindent Similarly, in the $\tau < \mathcal{I}$ case, we define the variables in Table~\ref{tab:cov4vars} such that the physical parameter covariance matrix is

\begin{equation}
\text{Cov}(\{t_c,b^2,\tau_0^2,r,f_0\},\{t_c,b^2,\tau_0^2,r,f_0\};~\tau < \mathcal{I}) =
\frac{\sigma^2}{\Gamma} \left(
\begin{array}{ccccc}
 \frac{3 \mathcal{I}}{b_{14} f_0^2 r^4} & 0 & 0 & 0 & 0 \\
 0 & \frac{B_1+B_5+B_6}{4 f_0^2 b_7 \mathcal{I} r^2} & \frac{\tau ^2 (B_1-B_5)}{16 f_0^2 b_7 \mathcal{I} r^4} & \frac{B_7}{4 f_0^2 b_7 \mathcal{I} r^2} & \frac{B_2}{2 f_0 b_7 \mathcal{I} r} \\
 0 & \frac{\tau ^2 (B_1-B_5)}{16 f_0^2 b_7 \mathcal{I} r^4} & \frac{\tau ^4 (B_1+B_5-B_6)}{64 f_0^2 b_7 \mathcal{I} r^6} & \frac{\tau ^2 B_8}{16 f_0^2 b_7 \mathcal{I} r^4} & \frac{\tau ^2 B_3}{8 f_0 b_7 \mathcal{I} r^3} \\
 0 & \frac{B_7}{4 f_0^2 b_7 \mathcal{I} r^2} & \frac{\tau ^2 B_8}{16 f_0^2 b_7 \mathcal{I} r^4} & \frac{B_9}{4 f_0^2 b_7 \mathcal{I} r^2} & \frac{B_4}{2 f_0 b_7 \mathcal{I} r} \\
 0 & \frac{B_2}{2 f_0 b_7 \mathcal{I} r} & \frac{\tau ^2 B_3}{8 f_0 b_7 \mathcal{I} r^3} & \frac{B_4}{2 f_0 b_7 \mathcal{I} r} & \frac{b_{10}}{b_7 \mathcal{I}} \\
\end{array}
\right).
\label{eqn:cov4}
\end{equation}

\bibliography{biblio}

\begin{thebibliography}{}
\expandafter\ifx\csname natexlab\endcsname\relax\def\natexlab#1{#1}\fi

\bibitem[{Caprio(2005)}]{Caprio2005CPC}
Caprio, M. 2005, Computer Physics Communications, 171, 107

\bibitem[{{Carter} {et~al.}(2008){Carter}, {Yee}, {Eastman}, {Gaudi}, \&
  {Winn}}]{CarterEt2008ApJ}
{Carter}, J.~A., {Yee}, J.~C., {Eastman}, J., {Gaudi}, B.~S., \& {Winn}, J.~N.
  2008, \apj, 689, 499

\bibitem[{{Claret} \& {Bloemen}(2011)}]{ClaretEt2011AA}
{Claret}, A., \& {Bloemen}, S. 2011, \aap, 529, A75

\bibitem[{{Coe}(2009)}]{Coe2009arXiv}
{Coe}, D. 2009, ArXiv e-prints, arXiv:0906.4123

\bibitem[{{Eastman} {et~al.}(2013){Eastman}, {Gaudi}, \&
  {Agol}}]{EastmanEt2013ASP}
{Eastman}, J., {Gaudi}, B.~S., \& {Agol}, E. 2013, \pasp, 125, 83

\bibitem[{{Foreman-Mackey} {et~al.}(2013){Foreman-Mackey}, {Hogg}, {Lang}, \&
  {Goodman}}]{ForemanMackeyEt2013PASP}
{Foreman-Mackey}, D., {Hogg}, D.~W., {Lang}, D., \& {Goodman}, J. 2013, \pasp,
  125, 306

\bibitem[{Goodman \& Weare(2010)}]{goodman2010ensemble}
Goodman, J., \& Weare, J. 2010, Communications in Applied Mathematics and
  Computational Science, 5, 65

\bibitem[{{Gould}(2003)}]{Gould2003}
{Gould}, A. 2003, ArXiv Astrophysics e-prints, astro-ph/0310577

\bibitem[{{Kipping}(2010)}]{Kipping2010MNRAS}
{Kipping}, D.~M. 2010, \mnras, 408, 1758

\bibitem[{{Kipping}(2013)}]{Kipping2013MNRAS}
---. 2013, \mnras, 435, 2152

\bibitem[{{Mandel} \& {Agol}(2002)}]{Mandel&Agol2002ApJ}
{Mandel}, K., \& {Agol}, E. 2002, \apjl, 580, L171

\bibitem[{{Seager} \& {Mall{\'e}n-Ornelas}(2003)}]{SeagerMO2003ApJ}
{Seager}, S., \& {Mall{\'e}n-Ornelas}, G. 2003, \apj, 585, 1038

\bibitem[{{Vallisneri}(2008)}]{Vallisneri2008}
{Vallisneri}, M. 2008, \prd, 77, 042001

\bibitem[{{Winn}(2011)}]{Winn2011}
{Winn}, J.~N. 2011, {Exoplanet Transits and Occultations}, ed. S.~{Seager},
  55--77

\end{thebibliography}

\clearpage
\begin{sidewaystable}

\caption{\label{tab:DerivativesF1} Partial derivatives of binned flux model the for $\tau > \mathcal{I}$ case.}
\begin{center}
\begin{tabular}{lccccc}
\hline\hline
& Totality & Totality/ingress/egress & Ingress/egress & Ingress/egress/out-of-transit & Out-of-transit \\
\hline
$\partial F_{lb1} \big/ \partial t_c$ & $0$ & $-\delta  (-T+\mathcal{I}+\tau +2 \left| t-t_c\right| ) ~\text{sgn}(t-t_c) / (2 \mathcal{I} \tau)$ & $-\delta  ~\text{sgn}(t-t_c) / \tau$ & $-\delta (T+\mathcal{I}+\tau -2 \left| t-t_c\right| ) ~\text{sgn}(t-t_c) / (2 \mathcal{I} \tau)$ & $0$ \\
$\partial F_{lb1} \big/ \partial \tau$ & $0$ & $-\delta  (-T+\mathcal{I}-\tau +2 \left| t-t_c\right| ) (-T+\mathcal{I}+\tau +2 \left| t-t_c\right| ) / (8 \mathcal{I} \tau ^2)$ & $\delta  (T-2 \left| t-t_c\right| ) / (2 \tau ^2)$ & $\delta  (T+\mathcal{I}-\tau -2 \left| t-t_c\right| ) (T+\mathcal{I}+\tau -2 \left| t-t_c\right| ) / (8 \mathcal{I} \tau ^2)$ & $0$ \\
 $\partial F_{lb1} \big/ \partial T$ & $0$ & $-\delta  (-T+\mathcal{I}+\tau +2 \left| t-t_c\right| ) / (4 \mathcal{I} \tau)$ & $-\delta  / (2 \tau)$ & $-\delta (T+\mathcal{I}+\tau -2 \left| t-t_c\right| ) / (4 \mathcal{I} \tau)$ & $0$ \\
$\partial F_{lb1} \big/ \partial \delta$ & $-1$ & $(-T+\mathcal{I}+\tau +2 \left| t-t_c\right| )^2 / (8 \mathcal{I} \tau) - 1$ & $-(T+\tau -2 \left| t-t_c\right|) / (2 \tau)$ & $-(T+\mathcal{I}+\tau -2 \left| t-t_c\right| )^2 / (8 \mathcal{I} \tau)$ & $0$ \\
$\partial F_{lb1} \big/ \partial f_0$ & $1$ & $1$ & $1$ & $1$ & $1$ \\
\hline
\end{tabular}
\end{center}

\caption{\label{tab:DerivativesF2} Partial derivatives of binned flux model for the $\tau < \mathcal{I}$ case.}
\begin{center}
\begin{tabular}{lccccc}
\hline\hline
& Totality & Totality/ingress/egress & Ingress/egress & Ingress/egress/out-of-transit & Out-of-transit \\
\hline
$\partial F_{lb2} \big/ \partial t_c$ & $0$ & $-\delta  (-T+\mathcal{I}+\tau +2 \left| t-t_c\right| ) ~\text{sgn}(t-t_c) / (2 \mathcal{I} \tau )$ & $-\delta  ~\text{sgn}(t-t_c) / \mathcal{I}$ & $-\delta  (T+\mathcal{I}+\tau -2 \left| t-t_c\right| ) ~\text{sgn}(t-t_c) / (2 \mathcal{I} \tau )$ & $0$ \\
$\partial F_{lb2} \big/ \partial \tau$ & $0$ & $-\delta  (-T+\mathcal{I}-\tau +2 \left| t-t_c\right| ) (-T+\mathcal{I}+\tau +2 \left| t-t_c\right| ) / (8 \mathcal{I} \tau ^2)$ & $0$ & $\delta  (T+\mathcal{I}-\tau -2 \left| t-t_c\right| ) (T+\mathcal{I}+\tau -2 \left| t-t_c\right| ) / (8 \mathcal{I} \tau ^2)$ & $0$ \\
$\partial F_{lb2} \big/ \partial T$ & $0$ & $-\delta  (-T+\mathcal{I}+\tau +2 \left| t-t_c\right| ) / (4 \mathcal{I} \tau )$ & $-\delta / (2 \mathcal{I})$ & $-\delta  (T+\mathcal{I}+\tau -2 \left| t-t_c\right| ) / (4 \mathcal{I} \tau )$ & $0$ \\
$\partial F_{lb2} \big/ \partial \delta$ & $-1$ & $(-T+\mathcal{I}+\tau +2 \left| t-t_c\right| )^2 / (8 \mathcal{I} \tau ) - 1$ & $-(T+\mathcal{I}-2 \left| t-t_c\right| ) / (2 \mathcal{I})$ & $-(T+\mathcal{I}+\tau -2 \left| t-t_c\right| )^2 / (8 \mathcal{I} \tau )$ & $0$ \\
$\partial F_{lb2} \big/ \partial f_0$ & $1$ & $1$ & $1$ & $1$ & $1$ \\
\hline
\end{tabular}
\end{center}
\end{sidewaystable}
\clearpage

\end{document}